\documentclass[12pt]{article}
\usepackage[utf8]{inputenc}
\usepackage[T1]{fontenc}
\usepackage[margin=1in]{geometry}
\usepackage{amsmath,amssymb}
\usepackage{booktabs}
\usepackage{longtable}
\usepackage{array}
\usepackage{natbib}
\usepackage[hidelinks]{hyperref}
\hypersetup{pdftitle={Seventy Years of Identifying the Phillips Curve and the Policy Rule},
  pdfauthor={Eric Vansteenberghe},
  pdfkeywords={identification, Phillips curve, policy rule, observational equivalence, monetary policy}}
\usepackage{setspace}
\newcommand{\lstar}{\lambda^{*}}

\newcommand{\ep}{\varepsilon}

\begin{document}

\begin{titlepage}
\begin{center}
{\Large\sc Seventy Years of Identifying\\[4pt]
the Phillips Curve and the Policy Rule}%
\\[18pt]
{\large Eric Vansteenberghe}\footnote{Banque de France and Universit\'{e}
Paris~1 Panth\'{e}on-Sorbonne. E-mail:
\href{mailto:eric.vansteenberghe@banque-france.fr}{\nolinkurl{eric.vansteenberghe@banque-france.fr}}.
The views expressed in this paper are solely those of the author and do
not necessarily reflect the views of past, present, or future employers.}
\\[10pt]
October 2026
\end{center}

\vspace{1.5em}
\noindent\textbf{Abstract.} Two coefficients enter nearly every quantitative argument about monetary policy: the Phillips-curve slope and the policy rule's response to inflation. For seventy years, each generation that estimated them identified the coefficient conditional on assumptions it maintained, and its successor relaxed one of them, naming the bias this exposed. This survey reconstructs the two programs as a single identification problem. It distinguishes five senses in which a coefficient has been called identified, reads each generation of the curve and of the rule in that key, and shows that the programs fail together because each borrows its identifying condition from the other. The same sequence ran on the latent inflation target and on the optimal rule. The survey closes on what the credible designs license when they leave a range rather than a point: bounds, signs and orderings, and decisions designed for a range.

\vspace{1em}
\thispagestyle{empty}
\end{titlepage}

\section{Introduction}
\label{sec:intro}

Two coefficients carry most of what a central bank asks empirical macroeconomics to deliver. The slope of the Phillips curve, $\kappa$, says how much disinflation costs; the inflation response of the policy rule, $\varphi_\pi$, says how much the bank has been willing to pay. Every quantitative answer to a question about monetary policy---how large the sacrifice ratio is, whether the Great Inflation was a policy failure, how far a bank may look through a supply shock---is built from the two, separately or as a composite. Each has been estimated for seventy years, and each estimate has been corrected by the next generation, which brought new data, new methods or new theory and named the bias the earlier estimate carried. This review reconstructs the two sequences side by side, from \citet{phillips1958relation} and \citet{taylor1993discretion} to the designs of 2026, and asks a question the corrections themselves never asked: in which sense was each coefficient identified, and why do the two programs fail together? %

The reconstruction yields three regularities. First, the corrections are re-attributions rather than refinements: what changes between generations is not the estimate but the sense in which it is an estimate of the named object---a fitted slope shown to be conditional on the expectations it assumed, a ``structural'' coefficient shown to be identified only within its model, a regime shift shown to be the persistence of a disturbance. Second, the two programs fail jointly, not independently: the curve's estimable slope depends on the rule that generated the data, flattening as the rule stabilizes inflation more successfully \citep{mcleay2020optimal}, and in the baseline New Keynesian model the rule's response coefficient is not identified from equilibrium data, because the determinacy the Taylor principle is meant to secure rests on responses off the equilibrium path, which the data never display \citep{cochrane2011determinacy}. Any object built from both inherits the two failures in correlated form. Third, each more credible generation narrows the object it identifies---from the aggregate price-side slope to a regional, conditional, often wage-side one \citep{hazell2022slope,beraja2019aggregate}; from the Fed's rule to the rule that markets and forecasters perceive \citep{hamilton2011estimating, bauer2024perceptions}. To make the first regularity precise, Section~\ref{sec:senses} distinguishes the senses of ``identification'' the two literatures have used: a fitted relation, the effect of an identified impulse, a coefficient identified within a maintained model, a policy-invariant primitive, a set. %

The question is not antiquarian. A policy calculation that needs both coefficients inherits what the two programs could and could not identify, and inherits it in the correlated form the second regularity describes. The companion paper \citep{vansteenberghe2026tolerable} studies one such calculation, the share of an inflation deviation it is optimal for a bank to tolerate in a given episode, and Section~\ref{sec:close} returns to it as the motivating case: the reconstruction sets a floor under what any estimator run on the same equilibrium record can deliver for it.

Each history has been written separately, and neither as an identification history. For the curve, \citet{humphrey1985early} and \citet{sleeman2011rushed} recount its origins; \citet{gordon2011history} reads the sequence after 1975 as a bifurcation between a backward-looking tradition ``better suited to explaining the postwar US inflation process'' (p.~10) and a forward-looking one; \citet{forder2014myth} argues that the early story itself---Phillips's paper as a breakthrough, the trade-off as a menu the profession believed in, the expectations critique as its refutation---is a later construction; the Boston Fed retrospective \citep{fuhrer2009understanding} reviews the fifty years since Phillips's paper; the critical surveys of \citet{mavroeidis2014empirical} and, through the formation of expectations, \citet{coibion2018formation} take the New Keynesian generation and its corrections as their subject, and \citet{furlanetto2024slope} make endogeneity, which stabilizing policy aggravates, the organizing fact, but without asking about the rule. For the rule, \citet{asso2007taylor} trace the intellectual history from Simons and Phillips to adoption; \citet{orphanides2003historical} reads the Federal Reserve of the 1920s and of the postwar decades through the rule ``as an organizing device'' (p.~983) and makes real-time data the correction; \citet{taylor1999historical} and \citet{kahn2012estimated} review the estimated rules as guides to policy; \citet{khoury1990federal} had already found the reaction-function estimates of the 1960s–80s ``not robust'' (p.~28). The surveys of identification in macroeconomics \citep{nakamura2018identification, stock2018identification, ramey2016macroeconomic} organize the field around identified shocks rather than around these coefficients, and the general taxonomies of the concept and of policy evaluation \citep{lewbel2019identification, heckman2010building} do not reach monetary economics. The research papers that connect the two objects each condition one side on the other: \citet{mcleay2020optimal} on a known rule, \citet{mavroeidis2010monetary} on a known curve, \citet{delnegro2020whats} and \citet{bergholt2023did} on a system assumed to identify both---the step \citet{canova2009back} warned about. The ancestor of the joint problem is \citet{sargent1976observational}, whose two reduced forms fit equally well under the rule in force and differ only in what survives a change of it; its policymaker's version is the learning literature in which the rule is the government's optimal response to the Phillips curve it has estimated \citep{sargent1999conquest, cogley2005conquest, primiceri2006why, sargent2006shocks}---Section~\ref{sec:joint} returns to it. What this review adds is the two sequences in one frame, and a column the histories lack: the sense in which each generation's coefficient was identified. %

Three things this review does not do. It estimates nothing: every number in it is a published estimate, quoted with the assumptions under which it was obtained. It does not take a side on whether the Phillips curve ``exists'' or has ``died'': the question it asks of each generation is what was identified, in which sense, and under what maintained assumptions---a question that has an answer whatever the slope's value. And it is not a history of the concept of identification, which has its own literature; it borrows the concept's modern statement once, in Section~\ref{sec:senses}, and applies it. Coverage is selective by design: the works cited are those that made an identification claim, corrected one, or defined what the claim meant.

The review proceeds as follows. Section~\ref{sec:senses} fixes the five senses of identification and the key the tables use. Sections~\ref{sec:curve} and~\ref{sec:rule} reconstruct the two programs generation by generation, each closing on the object its credible designs deliver. Section~\ref{sec:joint} reads the two together and shows where each borrows its identifying condition from the other. Sections~\ref{sec:target} and~\ref{sec:optimum} follow the same ratchet on two further objects, the latent inflation target and the optimal rule. Section~\ref{sec:close} states what the reconstruction licenses---a range, a sign and a decision rule---and returns to the motivating case. %

\section{What ``identified'' has meant}
\label{sec:senses}

Identification has one definition and many uses. The definition is Koopmans's: two structures are ``(observationally) equivalent (or indistinguishable) if the two conditional distributions of endogenous variables generated by [them] are identical for all possible values of the exogenous variables'' \citep[p.~133]{koopmans1949identification}, and a parameter is identified when it takes the same value in every admissible structure that generates the same distribution as the true one---in \citeauthor{lewbel2019identification}'s words, ``model parameters or features being uniquely determined from the observable population that generates the data'' \citep[p.~835]{lewbel2019identification}. Everything turns on the class of structures within which uniqueness is claimed, as Koopmans's own summary says: identifiable is what the distribution of the endogenous variables implies, ``given the model (which is accepted as valid)'' (p.~133). The two literatures this review reconstructs have claimed it within five different classes, and a reader who does not keep them apart will read the seventy years as a sequence of estimates that kept moving, rather than as what it was: a sequence of claims whose sense kept falling. %

\paragraph{One coefficient, five claims.} The classes are easiest to tell apart on a single number. That the slope of the Phillips curve is $\hat\kappa$ can mean that a regression of inflation on slack returns $\hat\kappa$ in the sample at hand (S1); that a monetary shock identified outside the equation, from a narrative record or a market surprise, moves inflation and slack in that ratio (S2); that, if the New Keynesian Phillips curve is the true model, the data admit no other value (S3); that $\hat\kappa$ is a property of price setting that would survive a change in the central bank's rule (S4); or that the data and the assumptions one is willing to defend bound the slope to a set that includes $\hat\kappa$ without singling it out (S5). The five readings answer different questions. A fit describes the sample; an identified impulse measures the effect of an intervention that took place; a coefficient identified within a model supports counterfactuals inside that model; a primitive supports the evaluation of a rule never tried; a set says what the defensible assumptions leave open. The same five readings apply to the rule's response to inflation, $\varphi_\pi$. A policy calculation---the cost of a disinflation, the right response to a supply shock---needs the fourth: for the slope, a value that would survive a change of rule; for the rule, its response off the equilibrium path. What the evidence reconstructed below delivers, generation after generation, is one of the other four, or less. %

\paragraph{S1---a fitted relation.} The weakest sense: a coefficient of a regression that fits the sample used, or of a rule calibrated to it, with no claim that it would survive a change in how the data were generated. \citet{phillips1958relation} qualified at once the policy figures he read off his curve---``these conclusions are of course tentative'' (p.~299)---and \citet{taylor1993discretion} claimed no more for a rule that ``need not be a mechanical formula'' (p.~198) and whose coefficients were chosen, not estimated. What such a claim is worth is a matter of fragility: ``an inference is not believable if it is fragile, if it can be reversed by minor changes in assumptions'' \citep[p.~38]{leamer1983lets}. %

\paragraph{S2---the effect of an identified impulse.} The sense the modern literature on shocks has made its own. What is identified is a response---``estimates of responses to identified structural shocks---what applied microeconomists would call `causal effects'\,'' \citep[p.~60]{nakamura2018identification}---obtained from ``external sources of as-if randomness'' \citep[p.~917]{stock2018identification}: a narrative shock, a high-frequency surprise, a regional difference---or from a restriction internal to a vector autoregression, such as a sign restriction. The claim is about the impulse and its exclusion, not about any equation. The same impulse can then be used as an instrument for an equation's coefficient---the regional design of \citet{hazell2022slope}, the shock sequences of \citet{barnichon2020}, the instrumental route of \citet{mcleay2020optimal}. That second use, which we call S2b to keep it apart from the response itself (S2a), does not identify the equation in a new sense: the coefficient it recovers is still that of a maintained equation, with the instrument's exogeneity to the equation's residual as the added assumption, and the passage from a regional or shock-specific response to an aggregate slope is a further step: ``typically, regional responses can only be translated into aggregate responses through the lens of a fully specified general equilibrium model'' \citep[p.~66]{nakamura2018identification}. %

\paragraph{S3---a coefficient identified within a maintained model.} The classical sense, and the one most of the two programs' ``structural'' estimates claim: a behavioral equation's coefficient identified by moment conditions or by a model's cross-equation restrictions, unique within the class of structures the model admits and conditional on the model being the true one. It comes in degrees: a parameter may be identified only locally \citep[p.~578]{rothenberg1971identification}, only up to a normalization, only by the prior \citep{delnegro2008forming}, or so weakly that the data barely inform it---``weak identification is pervasive in U.S. data'' \citep[p.~139]{mavroeidis2014empirical}---and, in a dynamic system, the classical rank conditions ``rest on the assumption that the reduced form parameters are identifiable'' \citep[p.~1997]{komunjer2011dynamic}, while \citet{canova2009back} found, in the mapping from the structural parameters to the solution, ``observational equivalence, partial and weak identification problems'' to be ``widespread'' (p.~431). The tables tag the three estimation routes where the distinction matters: S3-G by GMM or another limited-information method, S3-FI by full-information likelihood, S3-P by priors. %

\paragraph{S4---a policy-invariant primitive.} The sense policy evaluation requires, and the one \citet{lucas1976econometric} made the criterion: ``any change in policy will systematically alter the structure of econometric models'' (p.~41), so that only a parameter invariant to the change can answer a question about it. \citet{sargent1976observational} gave the sharpest statement of what the criterion excludes---two reduced forms that ``have identical residuals and so fit equally well'' (p.~635) and differ only in which survives a change of rule---and \citet{chahrour2018news} proved an observational equivalence on the expectations side, where news and noise representations, two writings of one structure, are indistinguishable ``even given an ideal dataset'' (p.~1703). S4 cannot be established from one regime. Its empirical form is invariance across the regimes actually observed---super exogeneity, which ``is refutable in the data for past changes'' in the process of the conditioning variables and for all changes ``must remain a conjecture until refuted'' \citep[p.~285]{engle1983exogeneity}---a claim we label S4-test and treat as weaker than S4. Its usable members are the combinations that a policy question needs and that are ``often much easier to identify than the individual parameters themselves'' \citep[p.~359]{heckman2010building}: Marschak's maxim, of which the sufficient-statistic approach \citep{chetty2009sufficient} is the modern instance, for policy changes near those observed. %

\paragraph{S5---the set.} Not a sixth object but a mode of reporting available for any of the others: when the data and the assumptions the analyst is willing to defend do not deliver a point, ``the sharp identified set … is the set of parameters that exhaust all the information'' \citep[p.~178]{tamer2010partial}, and the honest report is the set, on the principle that ``the credibility of inference decreases with the strength of the assumptions maintained'' \citep[p.~12]{manski2013public}. In the two programs S5 is where several S3 claims ended: the identification-robust confidence interval for the price-stickiness parameter of the New Keynesian Phillips curve is ``unbounded from above'' \citep[p.~294]{kleibergen2009weak}, and the post-1979 set for the rule's response covers the active and the passive region alike \citep{mavroeidis2010monetary}.

\paragraph{How the senses organize the review.} The senses are ordered by what a claim requires. Three of them form a chain: a fitted relation requires a sample (S1); a model-conditional coefficient requires the sample and a model (S3); a primitive requires both and invariance to a class of policy changes (S4). The identified impulse (S2) is not a rung of the chain but a branch beside it: it requires an exclusion restriction and no model, and it rejoins the chain when the impulse is used as an instrument for an equation (S2b), at which point the equation's model is maintained again. The set (S5) is a mode of reporting any of the four. The five sections that follow apply this key in turn: to the slope of the curve (Section~\ref{sec:curve}, Table~\ref{tab:idhistory_curve}); to the rule's response to inflation (Section~\ref{sec:rule}, Table~\ref{tab:idhistory_rule}); to the two read together (Section~\ref{sec:joint}, Table~\ref{tab:designs}); to the latent inflation target (Section~\ref{sec:target}, Table~\ref{tab:target}); and to the optimal rule (Section~\ref{sec:optimum}, Table~\ref{tab:rungs}). A row is a generation (a design in Table~\ref{tab:designs}, a rung of a ladder of optimal-policy calculations in Table~\ref{tab:rungs}), and its last column, written `claimed $\to$ shown', records the sense in which the row's coefficient was claimed to be identified and the sense the work that followed showed had been delivered. %

Take the second row of Table~\ref{tab:idhistory_curve}, `S4 as criterion; S3 in the tests $\to$ S3, conditional on adaptive expectations'. The generation of 1967--72 took a natural rate invariant to monetary policy as its criterion and tested it through the sum of the coefficients on lagged inflation; \citet{sargent1971note} showed that the test presumed the expectations process it needed, so that what had been estimated was a coefficient conditional on adaptive expectations. The corrections the tables record take three forms. A demotion moves the claim down the chain, as in that row, or as when a regime shift in the rule is read as the persistence of a disturbance. A set replaces the point by what the same assumptions bound: a ``structural'' slope whose identification-robust interval is unbounded. A conditioning keeps the sense and names the assumption the estimate stands on: a regional slope that is the aggregate slope only through a model; a rule's residual that is a shock only if the decomposition holds. In every form the estimate stays; the sense in which it is an estimate of the named object falls, or the condition under which it is one is written out. That is what the two histories have in common, and it is why the third regularity---that each more credible generation narrows the object it identifies---is a consequence rather than an accident: credibility is bought by moving down the chain or by naming the condition, and the objects that survive are local, conditional and on-path. %

\paragraph{The key and the taxonomies in use.} The key classifies claims, and it borrows its cells from taxonomies built for other purposes. The definition is Koopmans's and \citeauthor{lewbel2019identification}'s, and the zoo of the latter---point, local, generic and set identification, identification by functional form---supplies the qualifiers that appear inside S3 and the content of S5. S2 is the identified-moments and external-instruments program of \citet{nakamura2018identification} and \citet{stock2018identification}, which classifies designs by the source of their exogenous variation; the key does not reproduce that classification, because its object is the coefficient a design is asked to deliver, not the shock. The degrees of S3 are the failures cataloged for estimated models---observational equivalence, partial and weak identification \citep{canova2009back}---and its three routes follow the line \citet{mavroeidis2014empirical} draw through the evidence on the Phillips curve, between the limited-information estimation their survey covers and the full-information and Bayesian system estimation it sets aside. S4 is the Cowles notion of a policy-invariant parameter whose history \citet{heckman2000causal} traces and which \citet{heckman2010building} sets against the treatment-effect approach: in his terms, S2a answers the first policy-evaluation problem, the effect of an intervention that occurred, and S4 the third, the effect of one never observed. S4-test is super exogeneity \citep{engle1983exogeneity}. S5 is \citeauthor{manski2013public}'s and \citeauthor{tamer2010partial}'s. What none of these supplies, and what the two histories need, is the right-hand side of the tables' sense column: the sense a claim was later shown to have delivered. %

Whether the Lucas critique bites is itself an empirical question, and the two programs answer it differently. \citet{rudebusch2005assessing} found ``a modest empirical importance'' of the critique for monetary policy models (p.~247) and judged it ``unlikely that the inability to detect structural shifts in the reduced form stems from the low power of the statistical test'' (p.~270); \citet{linde2001testing} had given the opposite reading: the super-exogeneity test ``is not capable of detecting the relevance of the Lucas critique in practice in small samples'' (p.~986). \citet{estrella2003monetary}, testing forward- and backward-looking Phillips and IS curves across the policy breaks they date in the rule, found that ``some forward-looking models from the recent literature may be less stable than their better-fitting backward-looking counterparts'' (abstract). On the curve the critique is not tested but realized, on McLeay and Tenreyro's reading: the estimable slope changed because the rule changed, from a covariation the regression could read to one it could not \citep{mcleay2020optimal}---the same generation that could not detect the critique with super-exogeneity tests can see it in the flattening. On the rule the critique reaches its limit: the part of the rule that carries the policy stakes, the response off the equilibrium path, is not merely non-invariant but absent from the data \citep{cochrane2011determinacy}, so that no equilibrium time series, on either side of a change of regime, can refute a claim about it; ``observational equivalence is the same as nonidentification'' \citep[p.~488]{cochrane2023fiscal}. Section~\ref{sec:joint} states what follows for any object built from both. %

\section{The curve}
\label{sec:curve}

The curve began as a wage relation and the identification problem began as a translation. This section follows the slope through ten generations, from the fitted relation of 1958 to the designs of 2026, and asks of each the question of Section~\ref{sec:senses}: in which sense the slope was identified, under which maintained assumptions, and which of those assumptions the next generation relaxed, naming the bias the earlier estimate had carried. The euro area is treated in a block at the end, because its sequence is the same one run on a shorter sample with weaker instruments. The section closes on the object: which $\kappa$ each credible modern design delivers, and which one a policy calculation needs. %

\subsection{The fitted relation, 1958--1960}

\citet{phillips1958relation} fitted a curve to 1861--1913, found the data of the following decades close to it, the deflation policy of the 1920s included (p.~295), read off it the unemployment rate compatible with stable prices, and closed on a caveat: ``These conclusions are of course tentative'' (p.~299). In the key this is S1 stated as S1---a wage equation for one country, with the author's own caveat. \citet{lipsey1960relation} formalized the theory of the fitted relation, a micro reaction function in which wage change rises with excess demand for labor, unemployment serving as the observable proxy, its nonlinear link to excess demand generating the curvature, and aggregation over heterogeneous markets raising the curve and generating the loops; it is an S3 reading of an S1 fit, and it already contains the term the next generation would name, as a cost-of-living regressor in the money-wage equation (pp.~10, 26). \citet{samuelson1960analytical} moved the object from wages to prices and from description to policy, presented a ``menu of choice between different degrees of unemployment and price stability, as roughly estimated from last twenty-five years of American data'' (p.~192), and warned in the same pages: ``What we do in a policy way during the next few years might cause it to shift in a definite way'' (p.~193). The claim is S1 with the shifts stated; the stable exploitable menu that later readers attributed to the paper is not in it. \citet{forder2014myth} argues that this attribution is itself a later construction: ``The idea that they had been highly influential in promoting an inflationist view is a later invention'' (p.~5), and \citet{hoover2015genesis} shows that Samuelson and Solow ran no regression---the price curve is a translation of an implicit wage curve through productivity, offered as ``simply our best guesses'' \citep[p.~193]{samuelson1960analytical} by authors who had just written that the identifications needed to tell demand-pull from cost-push inflation ``may be quite impossible from the experience of macrodata that is available to us'' (p.~191). The first generation therefore claimed less than the history usually credits it with, and the first correction was of a reading rather than of an estimate. %

\subsection{Expectations and the natural rate, 1967--1976}

The correction that the history does record was theoretical, and it was made in the strongest sense of the key. \citet{friedman1968role} found in Phillips's analysis ``a basic defect---the failure to distinguish between nominal wages and real wages'' (p.~8)---the maintained assumption of stable price expectations named and relaxed. He defined the natural rate as ``the level that would be ground out by the Walrasian system of general equilibrium equations, provided there is imbedded in them the actual structural characteristics of the labor and commodity markets'' (p.~8): a primitive invariant to monetary policy, S4, with the slope reduced to a property of unanticipated inflation---``there is always a temporary trade-off between inflation and unemployment; there is no permanent trade-off'' (p.~11). \citet{phelps1967phillips} gave the same object its equilibrium definition, the rate ``at which the actual rate of inflation equals the expected rate of inflation so that the expected inflation rate remains unchanged'' (p.~255), with a unit coefficient on expected inflation (p.~261). %

The empirical implementation lowered the sense at once. The accelerationist tests regressed wage or price inflation on a distributed lag of past inflation and read the sum of the lag coefficients as the test statistic for the natural-rate hypothesis: \citet{gordon1970recent} found it ``0.45, very substantially less than 1'' (p.~18), so that the equations ``confirm the Phillips curve argument that a reduction in the unemployment rate can be traded off against a higher but ultimately stable rate of inflation'' (p.~11). The S4 claim was being tested as an S3 coefficient conditional on adaptive expectations. \citet{sargent1971note} showed that the test rested on an identifying restriction---``in order to identify $\alpha$ \ldots\ some extraneous information must be imposed'', almost always the unit sum of the lag weights (p.~722)---and so presumed the expectations process it needed: under rational expectations the cross-equation restrictions between the inflation process and the wage equation do not imply the sum-of-weights test, so a sum below one is consistent with the natural-rate hypothesis whenever inflation is mean-reverting. As \citet{hall2018short} put it in retrospect, ``a coefficient of less than one on a distributed lag with coefficients summing to one is the expected outcome in the likely case of mean reversion'' (p.~124). In the key the correction is the relaxation of one maintained assumption---the expectations process---and the demotion of the test from a test of the S4 claim to a cross-equation restriction. %

\citet{lucas1972expectations} then supplied the equilibrium account of why a fitted curve exists at all without offering a menu: ``the Phillips curve emerges not as an unexplained empirical fact, but as a central feature of the solution to a general equilibrium system'' (p.~122). Its cross-country test \citep{lucas1973some} made the slope a function of the variance of demand---``the higher the variance of demand, the more unfavorable are the terms of the Phillips tradeoff'' (p.~334)---which is an S4 claim for the primitives and, for the slope itself, a demonstration that it is regime-dependent. \citet{lucas1976econometric} generalized the demonstration into the criterion the rest of this review uses: ``any change in policy will systematically alter the structure of econometric models'' (p.~41), and, for the curve in particular, ``over a sample period during which $\pi$ and $\sigma^2$ remain roughly constant \ldots\ (13) will appear to the econometrician to describe a stable trade-off between inflation and real output'' (p.~38). The critique delivered no invariant slope; it delivered the statement that a fitted slope is conditional on the regime, and the next generation answered it by imposing the restriction the tests had been unable to establish. \citet{gordon1977can} wrote that ``ironically, the `natural rate hypothesis,' in the form of a coefficient of unity on price inflation, is vindicated by the revisions in the official data'' (p.~265), with expectations still ``a distributed lag of past changes in the personal consumption deflator'' (p.~265): the unit coefficient restored by a revision of the data and then imposed, as ``the sensible constraint that sums of coefficients of prices on wages and wages on prices equal unity'' (p.~276), and invariance checked on one episode, by asking whether a wage equation ``specified in 1971 and estimated for pre-1971 data'' could explain the wage change that followed (p.~255). The second generation ends where the first did, one sense lower than it began: an S4 criterion whose empirical form is an S3 coefficient conditional on the expectations model. %

\subsection{The NAIRU and the triangle model, 1977--2013}

The reconstruction was pragmatic. \citet{gordon1997time} regressed inflation ``on three basic determinants: inertia, demand and supply'' (p.~14) and defined the natural rate by the unit sum: ``if the sum of the coefficients on these lagged inflation values equals unity, then there is a `natural rate' of the demand variable \ldots\ consistent with a constant rate of inflation'' (pp.~14--15); the NAIRU became a ratio of coefficients, allowed to vary over time under a smoothness prior---an S3 object with the prior doing part of the work (S3-P)---and \citet{gordon2013phillips} presented the equation's stability through a sixteen-year post-sample simulation. The claim on the slope was S3; what the equation delivered, by the program's own account, was S1 (a stable fit) for the slope and S5 for the intercept. \citet{staiger1997nairu} showed that the intercept is imprecisely estimated: ``the 95 percent confidence interval for the current value of the NAIRU based on the GDP deflator is 4.3 percent to 7.3 percent'' (p.~34)---a precision problem for a ratio of coefficients, not a problem for the slope, which both sides kept: ``the empirical estimates confirm a clear, negatively sloped Phillips curve'' (p.~40). \citet[p.~23]{gordon1997time} answered with an economic rather than a statistical criterion for choosing among NAIRU series; \citet{staiger1997nairu} answered that ``debating over whether the NAIRU is 4.5, 5.5 or 6.5 percent does little to enlighten monetary policy'' (p.~47), since forecasts are insensitive to the point chosen. In the key: S5 against S3-P on the intercept, the slope untouched as S1. %

The other corrections of this generation aimed at the regressor and at the equation's payoff. \citet{orphanides2002unreliability} showed that real-time output gaps are revised by as much as the gaps themselves---errors-in-variables in the activity variable, so that ex post slopes claim a precision the policymaker never had. And the equation's practical test failed: \citet[p.~2]{atkeson2001} found that ``none of the NAIRU forecasts is more accurate than the naive forecast'' that next year's inflation equals last year's, a verdict on forecasting that was widely read as a verdict on the slope. It was not one---\citet{stock2007has} found, for GDP inflation through 2004, the coherence between activity and changes in inflation at business-cycle frequencies ``large and relative [\emph{sic}] stable across periods'' (p.~31), a stability the same authors later located, on data through 2019, in the components that ``tend to be relatively well-measured and domestically determined'', since ``as long as core PCE is used, the Phillips curve puzzle remains for all the slack measures'' \citep[pp.~393, 425]{stock2020slack}. Two decades passed before the sharpest statement of why: ``precisely because monetary policy can be used to offset the effect of such output gaps on inflation, their effect on inflation should not be visible in the data'' \citep[p.~206]{mcleay2020optimal}; unforecastable inflation, they note after earlier work, ``is a prediction of DSGE models in which policy makers respond aggressively to stabilize inflation'' (p.~202). Forecast failure is what a stable structural slope under a stabilizing rule looks like in equilibrium data. %

\subsection{The New Keynesian Phillips curve by GMM, and its weak-identification correction, 1995--2014}

The restoration re-founded the curve on optimizing price setters and estimated it by GMM. \citet{roberts1995new} showed that ``several of the New Keynesian models \ldots\ have a common formulation that is similar to the expectations-augmented Phillips curve of Friedman and Phelps'' (p.~975) and estimated it with survey expectations. \citet{gali1999inflation} replaced the output gap, whose coefficient came out ``negative and significant, which is at odds with the prediction of the theory'' (p.~207), with real marginal cost, and concluded: ``Real marginal costs are a significant and quantitatively important determinant of inflation. Backward-looking price setting, while statistically significant, is not quantitatively important. Thus, we conclude that the New Keynesian Phillips curve provides a good first approximation to the dynamics of inflation'' (p.~195). \citet{gali2001european} extended it to the euro area; \citet{christiano2005nominal} backed the parameters out of the response to an identified monetary shock, with the impulse response as the first-stage estimand. The claim was S3---a coefficient identified by moment conditions under Calvo pricing and rational expectations---and its structural parameters, the Calvo probability and the backward-looking share, were read as primitives by \citet{gali1999inflation} and \citet{gali2001european}, while \citet[p.~42]{christiano2005nominal} treated the Calvo model as ``a useful reduced form''. %

The correction was econometric, and it came in two steps that the key keeps apart. The first named a misspecification bias. \citet{rudd2005new} showed that, when the instruments include lags of inflation that the estimated equation omits, ``the estimated coefficient $\hat\omega_f$ will be positive, even though the `true' model does not include a forward-looking component'' (p.~1172); \citet{gali2005robustness} answered that ``these claims are incorrect'' (p.~1107), and \citet{rudd2007modeling} maintained that the over-identification tests have ``essentially no power in this context'' (FEDS WP 2005-66, p.~23, n.~27). The second step showed the two positions to be jointly admissible, which is the S5 verdict. \citet{mavroeidis2005identification} showed that ``problems of weak instruments may arise, depending on the properties of the `exogenous' variables'' (p.~421); \citet{kleibergen2009weak} made the case for identification-robust inference, under which forward-looking dynamics still ``dominate backward-looking dynamics'' but the sets are ``consistent both with no backward-looking dynamics and very substantial backward-looking behavior. Moreover, we cannot reject the null hypothesis that the coefficient on the labor share is zero'' (p.~294). The program's own survey stated the limit: ``the literature has reached a limit on how much can be learned about the New Keynesian Phillips curve from aggregate macroeconomic time series. New identification approaches and new datasets are needed to reach an empirical consensus'' \citep[p.~124]{mavroeidis2014empirical}. One further relaxation belongs here because it turns on the slope itself: under positive trend inflation the reduced-form slope on marginal cost falls and higher-order leads enter, so that omitting them ``would result in omitted-variable bias on the estimated coefficients, if the omitted terms are correlated with the regressors'' \citep[p.~694]{ascari2014macroeconomics}. The S3 slope is a function of the inflation target: S4 fails for the slope, and the candidates for invariance retreat to the pricing primitives behind it. The generation ends at S5---the honest report of what S3-G delivered---with the survey naming the remedy the next generations take: new identification approaches and new data. %

\subsection{The missing disinflation and the flattening, 2011--2020}

The Great Recession posed the curve a test it seemed to fail, and the fits of the 2010s restored it by re-measuring its regressors. \citet{ball2011inflation} found that ``a simple Phillips curve predicts a deflation that did not occur'' (NBER WP p.~3) and that ``the curve was flatter before 1973 and after the Volcker disinflation, when inflation was relatively low and stable'' (NBER WP p.~14); \citet{coibion2015phillips} recovered the relation by re-measuring the expectations that belong in it---``if firms' inflation expectations track those of households, then the missing disinflation can be explained by the rise in their inflation expectations between 2009 and 2011'' (p.~197)---and found ``no evidence of a decline in the slope of the Phillips curve over time'' (p.~219) on that measure. Each rescue relocated the object: same name, different regressor. Each was an S1 claim (a fit restored) or, for the survey-augmented equation, an S3 claim conditional on the expectations proxy chosen. The flattening was the other inference: \citet{blanchard2016phillips} found that ``the slope of the Phillips curve \ldots\ has substantially declined. But the decline dates back to the 1980s rather than to the crisis'' (p.~31); \citet{delnegro2020whats} posed the alternative exactly---``the observed cyclical disconnect between inflation and real activity might be the result of either a flat Phillips curve---the slope hypothesis---or a flat aggregate demand, generated by a forceful response of monetary policy to inflation'' (p.~304). They identified a demand impulse through the excess bond premium, estimated a DSGE, and concluded that ``there is strong support for the slope hypothesis: the slope of the Phillips curve has fallen very substantially after 1990, although it has not gone all the way to zero'' (p.~305). In the key their first result is S2a (the response to a proxy shock, a conditional slope), becoming S2b once a hybrid curve is fitted to the responses, and the second S3-FI (a DSGE posterior); both read the expectations regime through a maintained specification---survey expectations in a hybrid curve (pp.~331--332), a drifting target fitted to long-run expectations in the DSGE (p.~336)---which is what the two later corrections relax. \citet{bergholt2023did} find ``a relatively stable supply curve slope once shocks to supply have been properly accounted for. The demand curve, instead, has flattened substantially'' (p.~1)---the policy hypothesis given the larger weight, at the price of a maintained Taylor-rule block (p.~24). And the regional generation finds ``only a modest decline in the slope of the Phillips curve since the 1980s'' \citep[p.~1299]{hazell2022slope}, whereas a specification without time fixed effects yields a far steeper pre-1990 curve (p.~1305): what the aggregate fits read as a fall in the slope was mostly the removal of the movement of long-run expectations, an ``upward bias in the slope of the accelerationist Phillips curve'' in the words of \citet[p.~1]{jorgensen2024anchored}, who recover a stable structural slope once the degree of anchoring is allowed to change; \citet{bundick2023did} document the anchoring itself in the response of far-forward inflation compensation to news, which ``becomes statistically indistinguishable from zero'' after the numerical target (p.~1317). \citet{inoue2025has} then made the weak-instrument-robust inference of the previous generation time-varying and turned it on the flattening: on their reading the slope weakened from the early 1980s and had steepened again by the pandemic. %
Whether the anchoring moment belongs to the target is itself a maintained reading, which Section~\ref{sec:target} takes up \citep{malmendier2026seemingly}. %
 The generation's claims were S1, S2a and S3-FI; the sense shown is S1 conditional on the expectations regime---and the regularity that the whole review records appears here in its cleanest form: better identification did not move the estimate so much as change what the estimate was an estimate of. %

\subsection{Policy endogeneity, 2014--2024}

The bias every naive aggregate fit carries was named by \citet{mcleay2020optimal}, in the statement this review's companion relies on. Under the optimal targeting rule, with cost-push shocks the only disturbance, ``a naive OLS regression of inflation on the output gap \ldots\ will produce a negative parameter estimate, $\hat\gamma_1 = -1/6$, reflecting the targeting rule (eq.~[2]), rather than a consistent estimate of the positive slope of the Phillips curve'' (pp.~207--208); ``the identification problem is a simple case of simultaneity bias'' (p.~208). The better the policy, the worse the identification: a flat reduced-form curve is what a steep structural one looks like under a bank that offsets demand. The paper lists three routes out, and states what each maintains. Controls for cost-push shocks, in the tradition of \citet{gordon1982inflation}, maintain that the econometrician is ``successfully controlling for each and every tradeoff inducing shock affecting the economy'' (p.~201); instruments that ``purge their output gap data of any cost-push shocks, leaving only the demand variation'' maintain exogeneity and relevance, and ``these instruments may be too weak'' (p.~201); and regional variation maintains that ``at the regional level, the endogenous response of monetary policy to demand shocks is switched off'' (p.~215), with ``omitted variable bias unless the econometrician can control for the effect of regional inflation expectations'' (p.~218). All three sit under the premise that ``our framework is built under the assumption that the Phillips curve always holds (an assumption we later corroborate in the data)'' (p.~200): the routes are controls or S2b instruments for an S3 equation, and the equation is still the model's. \citet{fitzgerald2014stable} and \citet{fitzgerald2024regional} had reached the regional route from the same diagnosis---``if the NAIRU model is assumed to be true, regional data can be used to identify the structural relationship'' (\citeyear{fitzgerald2014stable}, abstract)---with the maintained assumption stated: ``a key assumption \ldots\ is that the parameters across states are the same'' (Staff Report 614, p.~24). What this generation shows is a negative result about S1 and S3 on aggregate data: the fitted slope is the policy-induced covariation, and the recovered slope is conditional on the maintained New Keynesian block. %

\subsection{The regional escape, 2019--2026}

\citet{hazell2022slope} made the regional route the design, and stated its object with a precision the review adopts as its model. A panel regression of regional inflation on unemployment identifies $\psi$, not $\kappa$: ``we show that panel data estimates of the regional Phillips curve by prior researchers are estimates of $\psi$ in equation~(3) as opposed to estimates of $\kappa$'' (p.~1306). Their own estimate is of $\kappa$, the coefficient of a forward-resolved regional equation for non-tradeable prices, estimated by two-sample two-stage least squares with lagged regional instruments or a tradeable-demand-spillovers instrument, which give the same value (p.~1331); ``changes in the long-run monetary regime are absorbed by time fixed effects when the regional Phillips curve is estimated using a panel data specification'' (p.~1304); and the aggregate slope is inferred, not estimated: ``we use a multiregion model to infer the slope of the aggregate Phillips curve from our regional estimates'' (p.~1299), since ``the slopes of the regional Phillips curve for nontradeables and the aggregate Phillips curve are the same in our model'' (p.~1318). The companion paper takes two results: the slope is small, and ``the sharp drop in core inflation in the early 1980s was mostly due to shifting expectations about long-run monetary policy as opposed to a steep Phillips curve'' (p.~1299). In the key the regional slope is S2b with the external instrument and S3-G with the lagged ones, the aggregate one S3, and the authors say so themselves; the survey of identification in macroeconomics by two of the same authors states the aggregation step as a general rule \citep[p.~66]{nakamura2018identification}. \citet{beraja2019aggregate} make the aggregation step the subject: ``making inferences about aggregate business cycles from regional variation alone is difficult because of economic channels and shocks that differ between regional and aggregate economies'' (p.~1789). The regional design identifies a slope with an external instrument, as \citet{barnichon2020} did with shock sequences, and it says which slope; its correction---that the regional object is not the aggregate one---is not a demotion of the claim but the claim itself. %

\subsection{Old shocks and external instruments, 2018--2023} %
\label{sec:curve-oldshocks}

The second modern design instruments the aggregate equation with the other program's residuals. \citet{barnichon2020} ``propose a new approach that consists in using sequences of independently identified structural shocks as instrumental variables'' and, estimating ``a Phillips curve using monetary shocks as instruments'', ``find that conventional methods substantially underestimate the slope of the Phillips curve'' (p.~2255)---an unemployment-gap coefficient of $-0.42$ over 1969--2007 (p.~2276) against the smaller absolute values of lagged-instrument GMM, the difference named as ``confounding with supply factors'' and ``attenuation'' (p.~2281). The design states its conditions: exogeneity, $E(\varepsilon^i_{t:t-H}u_t) = 0$, and full-rank relevance (p.~2262). It is S2b---S3 with an external instrument---and its maintained assumption is the one the residual-as-shock literature of Section~\ref{sec:rule} contests: that the monetary shock is a shock. \citet{nakamura2018high} argued that ``Fed announcements affect beliefs not only about monetary policy but also about other economic fundamentals'' (p.~1283); \citet{jarocinski2020deconstructing} decomposed the surprise into a policy and a ``central bank information shock'' (p.~3), and \citet{bauer2023alternative} showed that the regressions of forecast revisions on surprises behind the information effect ``are also consistent with a `Fed response to news' channel'' (p.~664). Each conditions the instrument further: the old-shocks slope is identified conditional on the shock's exogeneity, as its authors state (p.~2275)---for the narrative shocks, that policy did not respond to cost-push information beyond the Greenbook; for the 1990--2017 surprises, on which decomposition of the surprise is maintained---and conditional on the long-run expectations regime over a sample that spans the 1970s. \citet{barnichon2021phillips} then moved to a statistic that needs no equation, the Phillips multiplier---``a simple instrumental variable regression of cumulative inflation on cumulative unemployment using monetary shocks as instruments'' (p.~689)---and found that ``the decline in the trade-off is mostly due to the anchoring of inflation expectations'' (p.~689): S2a, converging with the regional generation on the time series while still implying a larger trade-off than OLS (p.~695). The two cleanest modern designs thus land on opposite sides of the conventional estimate because they identify different objects---the regional design a forward-resolved non-tradeable $\kappa$ with the expectations regime differenced out, the shock design an aggregate coefficient conditional on the shock's exclusion---and neither engages the other. %

\subsection{The micro anatomy, 2024--2025}

\citet{gagliardone2025anatomy} relocate the problem rather than remove it. Estimating the ``primitive form'' of the curve on Belgian firm data, with marginal cost as the forcing variable and a lagged firm-productivity instrument, they ``find a high slope for the cost-based Phillips curve, which contrasts with the low estimates of the conventional output gap--based formulation'' (p.~3941), and reconcile the two by decomposition: the output-gap slope $\kappa$ ``is the product of the marginal cost-based slope ($\lambda$) and the output elasticity of marginal cost ($\sigma_y$)'' (p.~3965), which ``suggests that the flat slope of the conventional NKPC reflects a weak link between the output gap and marginal cost over this sample period rather than a limited transmission of marginal cost fluctuations to inflation'' (p.~3944). In the key, $\lambda$ is a firm-level S3-G coefficient read as S4 through the pricing primitives behind it; $\kappa$ is not, because $\sigma_y$ is an aggregate elasticity valid over the sample, and it is $\sigma_y$ that reconciles the steep primitive with the small regional slope. The anatomy therefore confirms a small activity slope with different machinery, and locates the reduced form exactly where a policy calculation on activity needs it: in the mapping from slack to cost, which the authors themselves describe as sample-specific and as able to ``rise rapidly and substantially when the economy faces large shocks'' (p.~3972), shocks under which $\lambda$ also moves with the frequency of price adjustment. %

\subsection{The 2021--2026 surge and the nonlinearity}

The surge corrected the flattening consensus, and the correction is a fitted feature whose identification is disputed. \citet{benigno2023nonlinear} proposed ``a non-linear New Keynesian Phillips curve (Inv-L NK Phillips Curve)'' (abstract) with a kink at a vacancy-unemployment ratio of one---a cutoff ``not precisely estimated'' \citep[p.~85]{benigno2024slanted}---and took from \citet{mcleay2020optimal} the premise that ``if supply shocks are adequately controlled for, the Phillips curve can be empirically recovered'' (p.~10); their \citeyear{benigno2024slanted} version drew the L ``without any attempt to add controls or claim identification'' (p.~87). \citet{gudmundsson2024shifting} found that ``Phillips curves shifted up and steepened across advanced economies'' and showed in a two-sector model that the features ``emerge endogenously'' when unbalanced recoveries run against a goods-sector supply constraint (abstract)---a fitted slope that moves with the shock mix. \citet{harding2023understanding} obtained the steepening from a Kimball aggregator in an estimated model, ``a nonlinear Phillips curve that has a flat slope when inflationary pressures are subdued and steepens when inflationary pressures are elevated'' (p.~S101)---S3-P, the curvature identified by functional form and prior. \citet{beaudry2025fragility} found the evidence for the aggregate and regional kinks ``very fragile'' (abstract): at the metropolitan level the interaction term vanishes once time fixed effects are included, and at the aggregate level it survives only with expert forecasts, ``the difficulty of knowing how to properly control for inflation expectations'' (p.~24) being the crux---which measure of short-run expectations, experts' or households' and firms', belongs in the equation; \citet{beaudry2025dominant} found ``a quasi-flat Phillips curve, which was popular prior to the pandemic, still fits the post-2020 US data well'' (NBER WP 32322, abstract) once expectations are in the equation, and \citet{reis2026why} that ``the kink in the Phillips curve is still present, even though it is not as large'' (p.~491). The designs that identify a kink identify it for their own object: \citet{gitti2024nonlinearities} with a shift-share instrument across metropolitan areas, where the regional curve ``is flat in slack labor markets and significantly steepens when labor markets tighten'' (abstract) and aggregation is listed under future research, while \citet{cerrato2022inflation}, with the same instrument, find the regional slope much steeper from 2021 and infer the aggregate slope through a two-region model; \citet{bunn2025curvy} at the firm level, where ``using a model is crucial, as general equilibrium effects can affect how individual firm responses aggregate to the macro level'' (p.~26). \citet{kocherlakota2026concavity} shows that in time-dependent pricing models the curvature runs the other way---the curve is ``asymptotically horizontal for high levels of economic activity, and asymptotically vertical for low levels of economic activity'' (p.~113)---so that the sign of the curvature is a property of the maintained friction, not a feature the data settle. \citet{bernanke2023caused} traced most of the same inflation to ``shocks to prices given wages'' (p.~1), with a wage equation identified ``by assuming that wages respond to other variables with a lag of one quarter'' (p.~10) that ``predicts wage behavior after 2019 reasonably well'' (p.~14)---an S4-test on one episode of an S3 equation, which the kinked reading averages and the expectations reading absorbs; allowing the coefficient on $v/u$ to differ above one, they ``found no evidence for nonlinearity'' (p.~32), $v/u$ being ``itself convex'' in unemployment (p.~32, n.~27). Seventy years on, the curve in best repair is the one \citet{phillips1958relation} fitted: ``the wage Phillips curve is in better health than its price counterpart'' \citep[p.~302]{delnegro2020whats}, \citet{gali2011return} having shown a structural wage equation to ``account reasonably well for the comovement of wage inflation and the unemployment rate'' (p.~436)---which is to say that the 1960 translation from wages to prices, the move that created the policy object, is where identification is still being lost. %

\subsection{The euro area}

\citet{gali2001european} found that ``the NPC fits Euro area data very well, possibly better than U.S. data'' (p.~1237), with a stronger forward-looking component than in the United States---S3-G, with the Calvo and backward-looking parameters read as primitives; \citet{bardsen2004econometric} showed that ``the favourable evidence for NPCMs on euro-area data reported in earlier studies'' depends ``on specific choices made about estimation methodology'' (p.~671), and \citet{tillmann2009new} that the bootstrapped fit ``is consistent with both a well fitting and a failing model'' (p.~463)---the S5 verdict, reached by sensitivity analysis, the bootstrap and encompassing tests rather than by identification-robust sets, which euro-area work has computed for the Phillips multiplier \citep[p.~67]{eser2020phillips} and, for the slope, in the companion paper discussed below. The central banks' own assessments run from a fitted relation to the policy-endogeneity diagnosis. \citet{berson2018courbe} found the euro-area slope ``restée faible mais stable et significativement différente de zéro depuis la crise'' (p.~1); \citet{eser2020phillips} named the bias in that fit---temporary mark-up shocks ``shift the Phillips Curve and move the equilibrium output gap and inflation in opposite directions, which confounds their structural relation'' (p.~60), so that ``the causal impact between slack and inflation as revealed by the structural estimates is substantially higher than implied by the reduced-form estimates'' (p.~62)---and took both modern routes: country-level variation (p.~63) and high-frequency surprises (pp.~65--67). The regional design arrived with \citet{beschin2025slope}: 168 regions over 1999--2023, fixed-effects regressions checked by lagged-unemployment and shift-share instruments, and a slope ``relatively flat, but statistically significant'' (p.~1) that steepens with region fixed effects and falls as time, expectations and country-time fixed effects are added, until, saturated, it measures responses ``relative to other regions within the same country at a given point in time'' (p.~14); the regional nonlinearity ``effectively disappears when we control for time-varying country specific factors'' (p.~24), as it does in the metropolitan panel of \citet{beaudry2025fragility}. \citet{neri2026awakening} read the 2021--22 steepening as ``a transitory steepening of the Phillips curve, which had remained persistently flat throughout the preceding low-inflation period'' (abstract), operating through the frequency of price changes---a state-dependent S1. The kink is itself disputed on euro-area data: \citet{faber2026kinky}, on a panel of eighteen euro-area countries with time fixed effects, find the curve markedly steeper when labor markets are tight, while attributing most of the post-pandemic surge to factors other than tightness. %
 The euro-area sequence ends where the American one does, with one difference: the aggregate step from the regional object has no external instrument, and the companion paper's own euro-area exhibit---strong first stages and identification-robust sets unbounded on the slope---is the S5 report the aggregate data support \citep{vansteenberghe2026tolerable}.

\subsection{Which $\kappa$}

The designs that survive the weak-identification correction do not estimate one number; they estimate four objects, and each names the step it leaves to a model. The regional design delivers a forward-resolved regional non-tradeable slope, and its authors state that the aggregate slope is inferred through a multi-region model; the shock design delivers the coefficient of a maintained aggregate equation conditional on the shock's exclusion; the micro anatomy delivers a firm-level price-setting primitive and leaves the slack-to-cost mapping as a sample-specific elasticity; the impulse-response designs deliver a slope conditional on the shock, the horizon and the policy block the VAR maintains. What a composite policy calculation needs---the sacrifice ratio, the verdict on an episode, the optimal look-through of a supply shock \citep{vansteenberghe2026tolerable}---is the aggregate price-side slope on the activity variable the rule reads, over the horizon the rule acts on, invariant to the rule itself. No design above claims that object, and the program's own survey says why it cannot be claimed from aggregate time series \citep[p.~124]{mavroeidis2014empirical}. The current generation no longer claims one aggregate slope; it claims local, conditional, component-specific and often wage-side slopes, and treats the passage from these to the aggregate as an identification problem of its own---an S3 inference, whatever the design that produced the input. That is the yield of seventy years, and it is not nothing: the sign is settled, the order of magnitude in the inflation-targeting era is settled once the objects are aligned, and the reasons the aggregate fit is not the structural slope are known and named. What is not settled is the object a policy calculation needs, and the next section shows that the rule's history ends at the same place.

{\footnotesize
\setlength{\LTcapwidth}{\textwidth}
\begin{longtable}{>{\raggedright\arraybackslash}p{2.0cm}>{\raggedright\arraybackslash}p{4.0cm}>{\raggedright\arraybackslash}p{4.4cm}>{\raggedright\arraybackslash}p{1.8cm}>{\raggedright\arraybackslash}p{2.1cm}}
\caption{Seventy years of identification, I: the curve ($\kappa$)---what each generation identified, in which sense, and the bias the next generation named. Sense tags as in Section~\ref{sec:senses} (S1 a fitted relation · S2a an identified impulse · S2b an impulse as instrument · S3 a model-conditional coefficient, by GMM or another limited-information method (S3-G), full-information likelihood (S3-FI) or priors (S3-P) · S4 a policy-invariant primitive · S4-test invariance across observed regimes · S5 a set); `$\to$' reads `claimed $\to$ shown'. The table is deliberately coarse; the text gives the qualifications and the page-pinned quotations.}\label{tab:idhistory_curve}\\
\toprule
Generation & What the generation identified & What the next generation showed & Object & Sense claimed $\to$ shown \\
\midrule
\endfirsthead
\multicolumn{5}{l}{\emph{Table~\ref{tab:idhistory_curve}, continued}}\\
\toprule
Generation & What the generation identified & What the next generation showed & Object & Sense claimed $\to$ shown \\
\midrule
\endhead
\bottomrule
\endlastfoot
Fitted curve, 1958–60 & a wage–unemployment relation for the UK, ``tentative'' \citep{phillips1958relation}; a price-side ``menu'' that ``might \ldots{} shift'' \citep{samuelson1960analytical} & the omitted term is expected inflation: ``the failure to distinguish between nominal wages and real wages'' \citep{friedman1968role}; the inflationist reading of the menu is ``a later invention'' \citep{forder2014myth} & UK wage; US price by translation; contemporaneous & S1 $\to$ S1; the S4 reading imputed by later readers \\
\addlinespace[2pt]
Natural rate and accelerationist tests, 1967–72 & a natural rate invariant to monetary policy; the slope a property of unanticipated inflation \citep{friedman1968role,phelps1967phillips}; tested as the sum of lag coefficients \citep{gordon1970recent} & the test presumed the expectations process it needed \citep{sargent1971note}: a sum below one ``is the expected outcome in the likely case of mean reversion'' \citep{hall2018short} & US wage and price; lags & S4 as criterion; S3 in the tests $\to$ S3, conditional on adaptive expectations \\
\addlinespace[2pt]
The critique, 1972–77 & a fitted curve exists ``as a central feature of the solution to a general equilibrium system'' \citep{lucas1972expectations}; ``any change in policy will systematically alter the structure of econometric models'' \citep{lucas1976econometric} & no invariant slope delivered; the unit coefficient restored---``vindicated by the revisions in the official data'' \citep{gordon1977can} & US price; lags & S4 as criterion $\to$ S3, invariance checked on one episode \\
\addlinespace[2pt]
NAIRU / triangle, 1977–2013 & inertia, demand and supply; the NAIRU as a ratio of coefficients under a smoothness prior; post-sample stability \citep{gordon1997time,gordon2013phillips} & the NAIRU imprecisely estimated: ``4.3 percent to 7.3 percent'' \citep{staiger1997nairu}; real-time gaps unreliable \citep{orphanides2002unreliability}; forecasts no better than a naive benchmark \citep{atkeson2001}---not a verdict on the slope \citep{mcleay2020optimal} & US price; revised gaps & S3-P $\to$ S1 for the slope, S5 for the intercept \\
\addlinespace[2pt]
NKPC by GMM, 1995–2005 & a structural, forward-looking slope on marginal cost, ``a good first approximation'' \citep{gali1999inflation}, extended to the euro area \citep{gali2001european}; parameters backed out of an identified impulse \citep{christiano2005nominal} & omitted-variable bias through lagged-inflation instruments \citep{rudd2005new}; ``these claims are incorrect'' \citep{gali2005robustness}; tests with ``essentially no power'' \citep{rudd2007modeling} & US and EA price; forward-resolved; unit labor cost & S3-G; S3 backed out of an S2a impulse $\to$ S3-G contested \\
\addlinespace[2pt]
Weak identification, 2005–14 & identification-robust sets \citep{kleibergen2009weak}, after weak instruments were shown \citep{mavroeidis2005identification} & the sets include zero and span purely forward-looking and predominantly backward-looking price setting; ``the literature has reached a limit'' \citep{mavroeidis2014empirical}; the slope moves with trend inflation \citep{cogley2008trend,ascari2014macroeconomics} & US price; aggregate time series & S3 $\to$ S5; S4 fails for the slope \\
\addlinespace[2pt]
Missing disinflation, 2011–19 & the curve rescued by re-measured core and a regime-varying slope \citep{ball2011inflation} and by household expectations \citep{coibion2015phillips} & ``simultaneity bias'' \citep{mcleay2020optimal}; ``essentially no missing disinflation'' once the pre-sample slope is not maintained \citep{hazell2022slope} & US price; contemporaneous & S1; S3 on a proxy $\to$ S1, regime-dependent \\
\addlinespace[2pt]
Flattening, 2016–23 & the slope fell \citep{blanchard2016phillips}; ``strong support for the slope hypothesis'' from an impulse identified through the excess bond premium and a DSGE \citep{delnegro2020whats} & the supply slope ``relatively stable'', the demand curve flattened \citep{bergholt2023did}; ``only a modest decline'' once long-run expectations are differenced out \citep{hazell2022slope}; ``upward bias in the slope of the accelerationist Phillips curve'' \citep{jorgensen2024anchored}; a further, post-2012 reduced-form flattening traced to anchoring \citep{bundick2023did}; the anchoring moment also produced by learning from experience \citep{malmendier2026seemingly}; a time-varying robust path on which the slope weakens from the early 1980s and steepens again \citep{inoue2025has} & US price; impulse response & S1; S2a/S2b + S3-FI $\to$ S1, conditional on the expectations regime \\
\addlinespace[2pt]
Policy endogeneity, 2014–24 & in their model, OLS on aggregate data recovers ``the targeting rule \ldots{} rather than a consistent estimate of the positive slope'' \citep{mcleay2020optimal}; regional data identify the structural relation ``if the NAIRU model is assumed to be true'' \citep[abstract]{fitzgerald2014stable}, a route \citet{fitzgerald2024regional} pursue & each route maintains a restriction: every shock controlled, instrument exogeneity, policy ``switched off'' regionally, and ``the Phillips curve always holds'' \citep{mcleay2020optimal} & aggregate, then regional; price & a negative result on S1/S3; S2b or S3-G routes $\to$ S3, each conditional on a maintained restriction \\
\addlinespace[2pt]
Regional escape, 2019–26 & earlier panels' ``estimates of $\psi$ \ldots{} as opposed to estimates of $\kappa$''; $\kappa$ itself small; the 1980s disinflation as long-run expectations; the aggregate slope inferred through ``a multiregion model'' \citep{hazell2022slope} & aggregation is model-dependent, as \citet{nakamura2018identification} and \citet{beraja2019aggregate} had already stated; the regional slope moves across regimes and kinks \citep{cerrato2022inflation,gitti2024nonlinearities} & regional non-tradeable price; forward-resolved & S2b or S3-G regional $\to$ S3 aggregate, by the authors' statement \\
\addlinespace[2pt]
Old-shocks IV, 2018–23 & monetary shocks as instruments; ``conventional methods substantially underestimate the slope'' \citep{barnichon2020}; the Phillips multiplier \citep{barnichon2021phillips} & the slope is conditional on instrument exogeneity, as the authors state (no response to cost-push information beyond the Greenbook for the narrative shocks, the decomposition for the 1990--2017 surprises): information effects \citep{nakamura2018high,jarocinski2020deconstructing}, ``Fed response to news'' \citep{bauer2023alternative}; the two modern designs identify different objects and neither engages the other & aggregate price; forward-looking NKPC; 1969–2007 & S2b $\to$ S3; S2a for the multiplier $\to$ S2a; each conditional on the shock's exogeneity \\
\addlinespace[2pt]
Micro anatomy, 2024–25 & a steep cost-based primitive; $\kappa = \lambda\sigma_y$ \citep{gagliardone2025anatomy} & the slack-to-cost mapping $\sigma_y$ is sample-specific and reconciles the steep primitive with the small regional slope; $\lambda$ itself moves with the frequency of price adjustment under large shocks (the authors' qualification) & Belgian firms; price; marginal cost & S3-G at the firm level, read as S4 for $\lambda$ $\to$ $\kappa$: S3, not S4 \\
\addlinespace[2pt]
The surge and the kink, 2021–26 & a kink at $v/u = 1$ \citep{benigno2023nonlinear} or at full-employment unemployment \citep{benigno2024slanted}; steepening from sectoral constraints \citep{gudmundsson2024shifting} and from Kimball curvature \citep{harding2023understanding} & ``very fragile'' to time fixed effects and to the expectations control \citep{beaudry2025fragility}; ``still present, even though it is not as large'' \citep{reis2026why}; a kink identified for the firm and the region \citep{bunn2025curvy,gitti2024nonlinearities}, aggregation left to a model; concave in time-dependent models \citep{kocherlakota2026concavity} & US and advanced-economy price; $v/u$ or unemployment; 2021–23 & S1 (S3 by supply controls for the 2023 kink); S3-P $\to$ S1, fragile to the expectations control; curvature a property of the maintained friction \\
\addlinespace[2pt]
The wage curve, 2011–25 & a structural wage equation that fits \citep{gali2011return}; ``in better health than its price counterpart'' \citep{delnegro2020whats}; a linear $v/u$ wage equation identified by timing that ``predicts wage behavior after 2019 reasonably well'' \citep{bernanke2023caused} & the 1960 translation from wages to prices is where identification is still lost; linearity averages the kink \citep[p.~11]{benigno2023nonlinear}; the surge carried by short-run expectations rather than slack \citep{beaudry2025dominant} & US wage; $v/u$ & S3; S4-test on one episode $\to$ S3, conditional on linearity and on the expectations terms \\
\addlinespace[2pt]
Euro area, 2001–26 & NKPC by GMM ``fits Euro area data very well'' \citep{gali2001european}; a stable, significant, small slope \citep{berson2018courbe}; the regional slope by fixed effects, checked by shift-share \citep{beschin2025slope}; a transitory steepening \citep{neri2026awakening} & the fit ``depend[s] on specific choices made about estimation methodology'' \citep{bardsen2004econometric} and is ``consistent with both a well fitting and a failing model'' \citep{tillmann2009new}; mark-up-shock simultaneity, the causal impact of slack in the structural estimates ``substantially higher'' \citep{eser2020phillips}; the saturated regional slope is a within-country relative object \citep{beschin2025slope}, its kink disputed \citep{faber2026kinky}; the aggregate robust sets unbounded \citep{vansteenberghe2026tolerable} & EA price; aggregate, country and regional & S3-G; S1; S2b regional; S1 state-dependent $\to$ S5 for the aggregate slope \\
\addlinespace[2pt]
\end{longtable}
}

\section{The rule}
\label{sec:rule}

The rule's history is shorter than the curve's and ends further from its object. Where the curve's slope is inferred through a model from something the data identify, the rule's inflation response has a part---what the central bank would do at inflation paths it never lets occur---that equilibrium data do not contain, and the history is the discovery of that fact by fourteen routes. This section follows the coefficient from the reaction functions of the 1960s to the perceived and targeted rules of 2026, asking of each generation which sense it claimed and which the next one showed; the object column of Table~\ref{tab:idhistory_rule} records, for each, whose rule was estimated, what was recovered, whether the object lies on or off the equilibrium path, and on which data vintage. The ECB block follows, and the section closes on which $\varphi_\pi$ a determinacy or tolerance calculation needs. %

\subsection{Descriptive reaction functions, 1963--1990}

Reaction-function estimation is older than its emblem. \citet{dewald1963objective} regressed an indicator of policy on the authorities' goal variables with no claim to regime-invariant coefficients: S1, a description of the weights. What the program delivered by 1990 was less than that. \citet{khoury1990federal}, surveying the estimates, asked whether, twenty-five years on, the studies ``generally have reached similar conclusions'' or whether their ``results \ldots\ are robust with respect to more or less arbitrary differences in specifications'', and answered that ``unfortunately, as this chapter will show, neither of those conditions has been met'' (p.~27). The fitted relation did not survive re-specification: not even S1. %

\subsection{Calibration, 1993--1999}

\citet{taylor1993discretion} did not estimate. The response to inflation was set at 1.5---``the coefficients are round numbers that make for easy discussion'' (p.~202); the rule was offered as representative and expressly not as a mechanical formula (p.~197), and its evidential claim was its fit to actual policy in 1987--92 (p.~202). In the key this is S1 by calibration, with no identification claim to correct; it was the reception that made the benchmark a quasi-structural object, and \citet{asso2007taylor} quote the sentence in which Taylor himself drew the inference the reception would harden: ``the Fed policy has been conducted as if the Fed had been following a policy rule much like the one called for by recent research on policy rules'' (p.~25). \citet{taylor1999historical} then estimated the rule across eras and read the difference causally: the inflation coefficient below one in 1960--79 and above one in 1987--97 (p.~330), and the earlier policy ``was responsible for bringing on the Great Inflation of the 1970s'' (p.~339). The same chapter carries, in a footnote, the two remarks that the next twelve years would unpack: ``As explained above this equation is actually a reduced form of several structural equations, especially in the gold standard and Bretton Woods periods \ldots\ No attempt has been made to correct the estimates for serial correlation of the errors in the equation. I want to allow for the possibility that monetary policy mistakes are serially correlated in ways not necessarily described by simple time-series models. In fact, this serial correlation is very large, especially in the gold standard period when the equations fit very poorly'' (p.~329, n.~9). The claim is S1 with an S4-test read as a regime change; the author's own footnote qualifies the structural reading, for the gold-standard and Bretton Woods eras above all, and names the persistence of the error that Section~\ref{sec:rule-nonid} will show the regression recovers. \citet{judd1998taylor} estimated the rule chairman by chairman in partial-adjustment form and found the real-rate response to inflation insignificant under Burns and positive under Volcker and Greenspan (Exhibits~1--3), with the caveat that ``it would be preferable to use the original data that policymakers actually were looking at when decisions about the funds rate were being made'' (n.~8)---a footnote that goes on to cite the working-paper version of \citet{orphanides2001monetary}, with which the fifth generation begins. %

\subsection{GMM regimes, 1998--2000}

\citet{clarida1998monetary} attached macroeconomic history to the estimates: ``since 1979 each of the G3 central banks has pursued an implicit form of inflation targeting'', and ``these central banks have been forward looking: they respond to anticipated inflation as opposed to lagged inflation'' (p.~1033)---a rule for the target rate with smoothing, estimated by GMM with lagged variables as instruments, S3-G. \citet{clarida2000monetary} made the regime the result: the response 0.83 before 1979 and 2.15 after (p.~157), and ``since the estimate of $\beta$ is consistently less than unity for the pre-Volcker rule, self-fulfilling fluctuations are possible in this regime'' (p.~171). The instruments are lags of the funds rate, inflation, the output gap, commodity inflation, M2 growth and the spread (pp.~155--156), and the claim is S3-G, used as S4: the coefficient as the invariant that switched, and its position relative to one as what ``could be an important source of the shift in macroeconomic behavior'' (p.~148) between the 1970s and the 1980s. Three later generations showed what the claim was conditional on: the data vintage (Section~\ref{sec:rule-realtime}), the strength of the instruments after 1979 (Section~\ref{sec:rule-weakid}), and the equilibrium's containing the coefficient at all (Section~\ref{sec:rule-nonid}). %

\subsection{The determinacy estimand, 2003--2026}

\citet{lubik2004testing} made the indeterminacy claim itself the estimand: ``according to the estimated model, U.S. monetary policy post-1982 is consistent with determinacy, whereas the pre-Volcker policy is not'' (p.~190), a posterior weight on the two regions of a prototypical New Keynesian model, with the concession that the approach ``is potentially sensitive to model misspecification'' (p.~191). \citet{woodford2003interest} gave the coefficient its normative stakes: ``the Taylor principle \ldots\ the requirement that a sustained increase in inflation eventually result in an increase in nominal interest rates that is even larger in percentage points---turns out to be the critical condition'' for determinacy (p.~46). The claim is S3-FI with priors, S3-P, and the region is an object of the model; each relaxation of the model redrew it. \citet{davig2007generalized} showed that under Markov switching ``policy can satisfy the Taylor principle in the long run, even while deviating from it substantially for brief periods or modestly for prolonged periods'' (p.~607)---at \citeauthor{lubik2004testing}'s estimates, the passive 1970s determinate if the passive regime was expected to last less than about five years and the active regime to be sufficiently persistent (pp.~620--621)---and \citet{farmer2010generalizing} that this condition ``does not establish uniqueness of the fundamental equilibrium'' (abstract), the set depending on the equilibrium class. Over all bounded equilibria, \citet{barthelemy2019monetary} found the principle, regime by regime, ``neither a necessary nor a sufficient condition for determinacy'' (p.~355), strong departures from it ``not allowed'' (p.~355) and two active regimes able to leave the equilibrium indeterminate; on \citeauthor{lubik2004testing}'s estimates the 1970s were indeterminate only if agents expected the passive regime to last with a probability above 0.93--0.94 (p.~379), a belief the published estimates do not settle (p.~380). \citet{coibion2011monetary} relocated the boundary: ``while the Fed likely satisfied the Taylor principle before Volcker, the US economy was still subject to self-fulfilling fluctuations in the 1970s'' (abstract), because, for a rule responding only to current inflation, with trend inflation of six percent ``the central bank would have to raise interest rates by almost ten times the increase in the inflation rate to sustain a determinate REE'' (p.~347); and the pre-Volcker point estimates, whether below one as in \citeauthor{clarida2000monetary} or above one as in \citeauthor{orphanides2004monetary}, ``consistently come with such large standard errors'' that whether the economy was in a state of indeterminacy before Volcker ``remains unsettled'' (p.~341), although \citet{clarida2000monetary} report their own baseline coefficient as ``significantly below unity'', at $0.83$ with a standard error of $0.07$ (p.~157). \citet{hirose2020monetary} estimated the generalized model jointly and found ``substantial uncertainty about whether the policy response to inflation was active or passive during the Great Inflation'' (abstract), with a stronger response to output growth and a weaker one to the output gap substituting for the inflation response in the boundary. The last relaxation reversed the substitution: \citet{meyergohde2026informational} show that ``if the long-run Phillips curve is vertical, the Taylor principle is necessary and no amount of output-gap, output-growth, or forecast targeting can substitute for the vigorous response to inflation it prescribes'' (p.~1), the bounds being ``more conservative, placing tighter restrictions on monetary policy, and more robust, being independent of model-specific parameters and depending solely on monetary policy's reaction to current inflation'' (p.~1)---so that an interval straddling one cannot be rescued by the output responses on which the rescue relied, and the indeterminacy question is reinstated in the class of models with a vertical long-run curve. The estimand cannot be point-identified: it is reported as a set drawn by the maintained model, S5 conditional on the model. The verdict has proved model-relative: it moves with the pricing block and with beliefs about future regimes, not with the data on the rule. %

\subsection{Real-time data, 2001--2017}
\label{sec:rule-realtime}

\citet{orphanides2001monetary} re-ran the rules on the data policymakers saw: ``estimated policy reaction functions based on ex post revised data provide misleading descriptions of historical policy'' (p.~964). The result that carries the moral is in \citet{orphanides2004monetary}: on Greenbook forecasts the reaction to inflation forecasts was strong in both periods, and ``prior to Volcker's appointment, policy was too activist in reacting to perceived output gaps that retrospectively proved overambitious'' (p.~151)---the same rule, the same 1979 break, another vintage, the opposite verdict, and the 1970s relocated from the coefficient to the perceived gap. \citet{orphanides2003historical} made the contrast the method, distinguishing ``between real-time and retrospective renditions of history, as seen through the lens of Taylor rules'' (p.~994), and noted that ``in real time, gap-based forecasts of inflation are largely uninformative and often misleading'' (p.~1000, n.~24). The regressors' benchmarks are themselves estimates: real-time gaps carry revisions ``of the same order of magnitude as the estimated gap itself'', which ``are highly persistent'' \citep[p.~569]{orphanides2002unreliability}, and the rule's intercept stands on a natural rate that must be filtered from the data---\citet{holston2017measuring} found their estimates of it imprecise, and \citet{orphanides2003historical} that ``alternative assumptions regarding the equilibrium real interest rate would be required to reconcile the stance of a rule with either actual policy, or policy desired to achieve a specific operational objective for inflation on average'' (p.~996). In the key the real-time generation identified the perceived rule---the central bank's own reading of its information set---in sense S1 or S3-G conditional on that set, with the S4-test passing for the inflation coefficient and failing for the gap; and its own correction followed from the same data. \citet{boivin2006has} let the coefficients drift on real-time forecasts and found the response to inflation weak in the later 1970s, perhaps below one, but strong before 1973 and regaining strength from the early 1980s, naming the subsample average, more than the vintage, as the reason the estimates conflict---a chairman subsample is not a regime. %

\subsection{Weak identification, 2008--2011}
\label{sec:rule-weakid}

The critique that had reached the curve reached the rule. \citet{mavroeidis2010monetary} found that, although the point estimates imply determinacy, the identification-robust confidence sets also admit passive policy after 1979, so that ``the reaction function cannot be accurately estimated using data after 1979'' (p.~502): the pre-1979 passivity is relatively robust; the post-1979 activeness---the half of the story that carries the policy moral---is not. The first of the three explanations he offers (p.~502) is the regime itself: ``Good policy removes the possibility of sunspot dynamics, and mitigates the effects of shocks on future inflation and output. As a result, the expectations of these variables become less variable than they might otherwise be \ldots\ After 1979, sunspot dynamics may have subsided, which explains why the reaction function is poorly identified'' (p.~497). It is the rule-side twin of the curve's policy-endogeneity result: the better the policy, the less the data say about it---though, the author adds, the converse inference is barred: ``it is not possible to infer from the weak identification of the policy rule after 1979 that policy was in fact active'' (p.~492). \citet{inoue2011identifying} found a related masking in a New Keynesian model and a structural VAR: the sources of instability, the policy parameters among them, ``appear to have offsetting effects on output volatility'' (abstract) and so escaped earlier tests. In the key: S5, the honest report of what the GMM design contains after 1979. %

\subsection{Inertia or persistent shocks, 2002--2012}

The celebrated inertia of the estimated rule may not be inertia. \citet{rudebusch2002term} observed that a lag coefficient near 0.8 ``would imply a large amount of forecastable variation in interest rates at horizons of more than 3 months, which is contradicted by evidence from the term structure of interest rates. The illusion of monetary policy inertia evident in the estimated policy rules likely reflects the persistent shocks that central banks face'' (p.~1161)---an off-path check, using expectations data, on an on-path coefficient. \citet{english2003interpreting} nested the two hypotheses and found ``partial adjustment plays an important role'' (abstract) even with serially correlated errors; \citet{rudebusch2006monetary} answered that the nested estimate's own 95 percent interval for the share of partial adjustment ``stretches from 8 percent to 88 percent'' (p.~103), and that the term-structure evidence ``is definitive and indicates that the actual amount of policy inertia is quite low'' (p.~85). \citet{coibion2012why} kept the on-path specification, changed the data to real time, changed the instruments, and concluded that ``the evidence strongly favors the interest-smoothing explanation'' (p.~126), since markets may fail to forecast rate changes for reasons other than an absence of inertia---the perceived rule need not equal the Fed's. The two sides disagree on whether market expectations are a valid check on the Fed's coefficient. Inside the quarterly rule the first-order nested estimate does not decisively separate the two hypotheses (S5), which the smoothing side grants under the heading ``The Limited Informativeness of Nested Specifications'' \citep[p.~136]{coibion2012why}; points come from what that nesting leaves out---higher-order smoothing, which the same side reads as favoring inertia (pp.~140--141), expectations data, or identified non-monetary shocks used as instruments, under which ``all of the estimates of interest smoothing are high'' (p.~129). %

\subsection{The residual as a shock, 1998--2023}

A parallel program corrected what the rule's residual was taken to be. \citet{christiano1999monetary} stated the recursiveness assumption---``the time $t$ variables in the Fed's information set do not respond to time $t$ realizations of the monetary policy shock''---which, with a linear feedback rule, ``justifies estimating policy shocks by the fitted residuals in the ordinary least squares regression of the Fed's policy instrument on the variables in the Fed's information set'' (p.~68), and conceded that the literature ``is silent'' on what the systematic component does to output and prices (p.~70, n.~4): S2a for the residual, S1 at most for the rule. \citet{rudebusch1998do} found the VAR reaction functions ``implausible and misspecified'' (p.~908) and their residuals barely correlated with the funds-rate shocks priced in futures markets (p.~923). \citet{romer2004new} purged the intended funds rate of the Greenbook forecasts ``to derive a measure free of systematic responses to information about future developments'' (p.~1055): S2a conditional on the Greenbook spanning the Fed's information. The high-frequency generation then found the announcement surprise itself to be composite: \citet{nakamura2018high} that ``forecasts about output growth also increase---the opposite of what standard models imply about a monetary tightening'' (p.~1283), so that announcements move beliefs about fundamentals as well as policy; \citet{jarocinski2020deconstructing} that ``ignoring the central bank information shocks biases the inference on monetary policy nonneutrality'' (p.~1); \citet{mirandaagrippino2021transmission} that a rate hike may reveal either a departure from the rule or stronger fundamentals, so that the instrument must be orthogonalized to the bank's projections and past surprises; and \citet{bauer2023alternative} that the same regressions ``are also consistent with a `Fed response to news' channel'' (p.~664), the surprise being predictable from public data---an ``omitted variables bias'' (p.~667) rather than an information effect. The object left over after fitting a rule is as contested as the coefficients; and the contest matters twice, because the curve's cleanest modern instrument (Section~\ref{sec:curve-oldshocks}) is this residual. %

\subsection{The system level, 2004--2011}

Moving the rule inside an estimated general-equilibrium system relocated the problem. \citet{smets2007shocks} estimated the rule jointly with the model's frictions and priors and found ``the mean of the long-run reaction coefficient to inflation \ldots\ relatively high (2.0)'' (p.~594), while reporting that ``three other parameters are clearly not identified'' and fixing them (p.~592): S3-P, the prior supplying what the likelihood lacks. The identification analysis that followed placed the rule's parameters where they belong. \citet{iskrev2010local} gave as his example of a parameter that does not affect the solution ``the unidentifiability of the Taylor rule coefficients in a simple New Keynesian model pointed out in Cochrane (2007)'' (p.~193); \citet{komunjer2011dynamic} found that one of the rule's three parameters ``needs to be further restricted'', a result that ``supports the argument of Cochrane (2011) that parameters of the Taylor rule are not identified without restrictions'' (p.~2018); \citet{canova2009back} that ``observational equivalence, partial and weak identification problems are widespread and typically produced by an ill-behaved mapping between the structural parameters and the coefficients of the solution'' (p.~431). \citet{delnegro2004priors} had already made the prior's weight itself an estimated quantity, and \citet{delnegro2008forming}, eliciting the prior from beliefs about steady states and second moments, found that ``standard macro time series cannot discriminate among theories that differ in the quantitative importance of nominal frictions'' (NBER WP 13741, abstract). The rule coefficient in an estimated DSGE is an S3-P point; in the small models it lies on a ridge the likelihood alone does not resolve, while the medium-scale model passes the local rank condition once two curvature parameters are fixed \citep[p.~199]{iskrev2010local}, and there the posterior of the inflation response has mean $2.04$ and a 90 percent band, $1.74$ to $2.33$, that excludes the prior mean of $1.50$ \citep[Table~1A, p.~593]{smets2007shocks}. %

\subsection{Drifting and switching coefficients, 2001--2011}

Nor do the coefficients sit still. \citet{cogley2001evolving} read a drifting-coefficient VAR as a drifting rule---``For the remainder of the 1970s, the rule was decidedly passive \ldots\ Monetary policy started becoming activist in 1981'' (p.~360)---and, answering the discussants who read the drift as volatility, \citet{cogley2005drifts} ``continue to find evidence that the VAR coefficients have drifted, mainly along one important direction'' (p.~293) while finding drifting variances too. \citet{primiceri2005time} found the systematic responses growing more aggressive, a change that ``has had a negligible effect on the rest of the economy'' (abstract), the long-run response to inflation remaining above one throughout---a counterfactual, off-path check that assigns the coefficient's movement little weight. \citet{boivin2006has}, as above, found the weak response confined to the later 1970s. \citet{simszha2006were} found that ``the best fit allows time variation in disturbance variances only. With coefficients allowed to change, the best fit is with change only in the monetary policy rule \ldots\ But the differences among regimes are not large enough to account for the rise, then decline, in inflation of the 1970s and 1980s'' (p.~54); \citet{benati2009var} showed the two readings to be observationally close in VAR moments---a model whose only change is the move from passive to active policy generates smaller innovation variances, minor policy counterfactuals and stable impulse responses, so that ``existing VAR evidence is also compatible with the `good policy' hypothesis'' (abstract). \citet{ang2011monetary} priced the coefficient in the term structure, ``from close to zero in 2003 to a high of 2.4 in 1983'' (abstract)---the Fed's rule as read from bond prices, investors being assumed to know its law of motion. In the key: S1 time-varying with a policy label, and S5 at the system level, since coefficient regimes and variance regimes need not be separable by the VAR. %

\subsection{Non-identification, 2006--2021}
\label{sec:rule-nonid}

The deepest cut is theoretical, and this review's companion relies on it. \citet{cochrane2011determinacy} argued that in the baseline New Keynesian model the inflation coefficient is not identified from equilibrium data even in principle: ``the Taylor rule is not identified without unrealistic assumptions. Thus, Taylor rule regressions do not show that the Fed moved from `passive' to `active' policy in 1980'' (p.~565); in the model's simplest version, a Fisher equation and the rule, ``a Taylor-rule regression of $i_t$ on $\pi_t$ will estimate the disturbance serial correlation parameter $\rho$ rather than the Taylor rule parameter $\varphi$''; ``$\varphi$ is not identified from data on $\{i_t, \pi_t\}$ in the equilibrium of this model. Inflation is supposed to jump to the one value for which accelerating inflation at rate $\varphi$ is not observed. If inflation does jump, there is no way to measure how fast the inflation would accelerate if it did not jump'' (p.~573). What the Taylor principle relies on is a threat of action off the equilibrium path, which an equilibrium history cannot, by construction, display: ``Taylor determinacy depends entirely on what the Fed would do away from the [$\pi^*$] equilibrium, which we can never see from data in that equilibrium'' (p.~594). What would identify the coefficient is stated as two assumptions---that the Fed's reaction to a deviation from the desired equilibrium is the same as the relation between equilibrium rates and equilibrium inflation, and an autoregressive equilibrium representation in the determinate zone with an error orthogonal to usable instruments---and \citet{cochrane2023fiscal} compresses the result: ``observational equivalence is the same as nonidentification: without additional identifying assumptions, the parameters that separate regimes such as $\gamma$ and $\varphi$ are not identified from time series of observable equilibrium variables'' (p.~488). In the key this is an S4-level negative---the coefficient of interest is the off-equilibrium response, and equilibrium regressions recover at most the residual's persistence (S1), which is what Taylor's footnote had called ``very large''. The replies do not dispute the logic; they change the maintained model. \citet{sims2008identification} showed that ``Cochrane's finding of non-identification is not a generic implication of the model, but is rather the result of a particular assumption on the policy rule'' (abstract), identification returning under a different assumption on the rule; \citet{carrillo2008comment} that non-identification ``holds only for economies that do not display inflation inertia or habit formation'' (abstract), identification borrowed from the private block's persistence; and \citet{backus2020identifying} that in macro-finance models ``identification of the rule's parameters requires restrictions on the form of the shock'' (abstract). \citet{carvalho2021taylor} relocated the question to the size of the bias, which they argue is small because policy shocks account for little of the regressors' variance, with the concession, in a footnote, that the sign ``depends on the assumption that $\phi_\pi > 1$'' (p.~142, n.~2)---so that the bias result cannot itself establish determinacy. Each reply restores S3 by adding an assumption---a rule form, a private-sector inertia, a shock restriction, a small policy-shock variance under determinacy---which is Cochrane's ``unrealistic assumptions'' seen from the other side. \citet{gali2025commentary} restates from inside the New Keynesian camp (Section~\ref{sec:rule-targeted}) the off-path point that the inflation-targeting literature had written into implementation as ``an out-of-equilibrium commitment'': an instrument rule whose response coefficient leaves the optimal path unchanged ``for any value of the scalar'', while ``the determinacy properties \ldots\ may of course depend on'' it \citep[p.~1265 and n.~34]{svensson2011inflation}. %

\subsection{The lower bound, 2013--2016}

After 2008 the instrument itself stopped being observed. \citet{wu2016measuring} used the shadow rate as the policy rate from 2009, arguing that once the lower bound binds, the shadow rate summarizes the stance of monetary policy, which the short rate, held at the bound, no longer does; \citet{krippner2013measuring} had proposed the same object as ``a `shadow short rate'\,'' (p.~135). \citet{krippner2020note} then showed that ``SSR estimates can be sensitive to minor choices in their estimation. Used subsequently in a small macroeconomic model, those sensitivities lead to wide variations in the inferred effects \ldots\ Therefore, it should not be presumed that any SSR series will necessarily be quantitatively useful'' (p.~951). In the key the shadow rate is S1, a measurement of the rule's left-hand variable by a model-constructed generated regressor, and any rule estimated on it inherits the construction's fragility. %

\subsection{Perceived rules, 2011--2026}

The frontier moved onto ground the tolerance question occupies. \citet{hamilton2011estimating} estimated, from news' effects on forecasts of conditions and of policy, the rule agents use to form expectations, and found that ``the market-perceived Federal Reserve policy rule changed: the output response vanished, and the inflation response path became more gradual but larger in long-run magnitude'' (p.~1)---identified news impulses instrumenting the perceived, not the Fed's, rule. \citet{bauer2024perceptions} estimate the perceived rule from a panel of professional forecasts and find its dependence on economic conditions varying over the policy cycle and updated after high-frequency policy surprises, suggesting that forecasters ``have imperfect information about the rule'' (p.~2227); they are explicit about the object: ``importantly, the perceived policy framework may differ from the Fed's historical behavior'' (p.~2228), the divergence being particularly pronounced under strong forward guidance (p.~2229). In the tightening of 2022 the perceived inflation coefficient itself rose after liftoff, and the timing suggests that forecasters and markets learned the rule from the Federal Reserve's rate increases \citep{bauer2024changing}. %
 \citet{sastry2026disagreement} separates the mechanisms---asymmetric information, ``different perceptions of the policy rule, and different confidence in public signals''---and finds that ``markets slightly misspecify the monetary rule and are significantly underconfident in public information'', with ``central bank private information and `information effects' \ldots\ quantitatively negligible'' (p.~486); \citet{bocola2024bond} read from bond markets a comovement ``positive and stable for over twenty years'' that ``has weakened substantially over the 2020-2022 period'', ``indicative of a monetary authority that places less weight on inflation stabilization'' (abstract). What these designs identify is the mapping from forecasts of conditions to forecasts of rates---a coefficient of beliefs, observable off the realized path because forecasters' conditional expectations vary across the cross-section and across news. It is the nearest thing in the literature to the off-path object of Section~\ref{sec:rule-nonid}, and its authors, \citet{bocola2024bond} excepted, say it is not the Fed's: S1 in forecasts, S2a in its updating, S3 within a belief model---the ancestor of the perceived-rule band that \citet{vansteenberghe2026tolerable} reads off dealers' scenario surveys. %

\subsection{Source-dependent and targeted rules, 2024--2026}
\label{sec:rule-targeted}

The post-2021 reassessments read policy as source-dependent. \citet{hofmann2026targeted} introduce ``targeted Taylor rules: monetary policy rules which allow for different reactions to demand– and supply–driven inflation'' and estimate that ``U.S. monetary policy has reacted nearly fourfold more strongly to demand– than to supply–driven inflation since Paul Volcker's Chairmanship'' (p.~1). The design's identification is inherited twice: the rule is estimated by OLS---``we follow closely [\citeauthor{carvalho2021taylor}]'' (p.~10)---with the endogeneity bias signed rather than removed (``the estimated coefficient of demand-driven inflation will likely be biased downwards \ldots\ while that the estimated coefficient of supply-driven inflation is expected to be biased upwards'', p.~10, n.~14), and the regressors are themselves identified by a maintained decomposition: ``inflation is demand–driven in a given sector if prices and quantities move in the same direction in that specific area of consumption'' (p.~12, n.~17), after \citet{shapiro2022decomposing}, or, in their 2026 version, a language model's reading of FOMC transcripts. \citet{nakamura2025beyond} ask ``the degree to which the Taylor rule is descriptive, and the degree to which it should be viewed as prescriptive'' (p.~2) and find that ``while the original Taylor rule (generally) fits well during the Greenspan era (1987-2006), it (generally) fits poorly during the preceding Volcker era (1980-1987) and since the early 2000s'' (p.~2)---S1 fits by era, an S4-test failed. \citet{gali2025commentary}, discussing them, states the object exactly: ``whenever $\phi_\pi > 1$, the previous rule guarantees that $\pi_t = \pi_t^*$ for all $t$ so that the second term on the right hand side will be zero in equilibrium, and there is no reason in principle why $i_t^*$ should even comove positively with $\pi_t^*$ \ldots\ the Taylor principle in this case applies only to off-equilibrium paths that are not observed by an econometrician'' (p.~7), so that ``such a mechanical assessment of monetary policy based on a Taylor rule may be highly misleading'' (p.~9). The frontier is thus the tolerance margin measured on two faces---the source-dependence of the response and the public's perception of it---and the off-path response the Taylor principle is about remains, by the statements of Cochrane and Galí alike, outside the equilibrium data. %

\subsection{The euro area and the ECB}

The ECB block repeats the sequence on a shorter sample. \citet{gerlach2000taylor} found that average EMU interest rates in 1990--98 ``moved very closely in relation to average output gaps and inflation as suggested by the Taylor rule'' (abstract), the exchange-market turmoil of 1992--93 excepted---S1 with Taylor's coefficients; \citet{gerdesmeier2003empirical} estimated a forward-looking rule for the synthetic euro area by GMM with lagged instruments---S3-G, without the weak-identification question; \citet{sauer2007using} found that contemporaneous rules make the ECB look accommodating and destabilizing, an impression that ``seems to be largely due to the lack of a forward-looking perspective in such specifications'' (abstract)---the sign of the Taylor-principle verdict specification-dependent; \citet{belke2011does} that real-time data raise the inflation and output-gap coefficients and real-time forecasts lower the output one---the coefficient moving with the information set. \citet{orphanides2013complexity} interpreted ECB policy with a calibrated difference rule on survey forecasts, whose outcome-based version they found robust across eleven models---S1 as a benchmark, S5 for robustness---and \citet{bletzinger2017lower} estimated it on survey forecasts, found the inflation coefficient ``effectively the same as the 0.5 value'' (p.~124), and read the residual from 2015 as a ``lower for longer'' stance; \citet{gerlach2014ecb} detected ``a shift after Lehman Brothers failed in September 2008'' and ``a shift back in the second half of 2010'' (p.~137)---an S4-test failed, the shift read as lower-bound behavior rather than a change in the primitive. %
\citet{tatar2025policy} set simple outcome-based rules, Taylor's and the difference rule above, against the surge: those responding to recent core or domestic inflation ``called for raising interest rates in 2021---well ahead of the rate hikes of the ECB'' (abstract), S1 by calibration. %
\citet{rostagno2021monetary} read the same history as ``a tale of `two regimes'\,'', in which the price-stability definition ``with a hard 2\% ceiling, functioned as a key shock-absorber in the relatively high-inflation years prior to the crisis, but offered a softer defence in the face of the disinflationary forces'' (ECB WP 2346, abstract), because expectations settled back whenever inflation threatened to cross the ceiling, agents counting on a forceful response from the ECB---an S4 claim argued from expectations' behavior, the authors' own augmented Taylor rule giving ``equal support for two distinct characterizations of the ECB's policy rule'' \citep[p.~150]{rostagno2021monetary}. Against all of these stands the stated rule. The Governing Council's strategy of 2003 aimed ``to maintain inflation rates close to 2\% over the medium term'' \citep{ecb2003strategy}; that of 2021 made the commitment ``symmetric'', requiring near the lower bound ``especially forceful or persistent monetary policy measures to avoid negative deviations from the inflation target becoming entrenched'' \citep{ecb2021strategy}; that of 2025 recognizes ``the importance of appropriately forceful or persistent monetary policy action in response to large, sustained deviations of inflation from the target in either direction, to avoid deviations becoming entrenched through de-anchored inflation expectations'', allowing ``for inevitable short-term deviations of inflation from the target'' \citep{ecb2025strategy}. The stated rule is an S4 clause by construction---a state-contingent response to large and sustained deviations, meant to keep entrenchment off the equilibrium path. The estimated rules are S1 and S3-G on-path relations whose inflation coefficient runs from about one-half for the change in the rate (difference rules) to about two (GMM) and above four in a smooth-transition rule \citep[p.~147]{gerlach2014ecb}, shifts with the vintage and the expectations measure, and switches smoothly in late 2008 and back in 2010; and no estimate tests what the clause asserts off the path. The companion paper's euro-area regime dating is the exhibit of what the on-path data show \citep{vansteenberghe2026tolerable}. %

\subsection{Which $\varphi_\pi$}

The designs that survive the sequence deliver six objects, and each names what it does not identify. An information-set-conditional GMM coefficient, which moves with the vintage, the constancy of coefficients and the instruments' strength, and whose post-1979 value cannot be point-identified. A posterior region conditional on a model and its priors, which moves with trend inflation, the equilibrium class and the pricing block, and which, in the determinate region, the likelihood does not contain. An identification-robust set. A residual and a stance---conditional on what the rule's error contains, and on a term-structure model. A perceived rule: the mapping from forecasts of conditions to forecasts of rates, the one object in this history the data reach off the realized path, and one its authors say may differ from the central bank's. And a source-dependent response whose regressors are maintained decompositions. What a determinacy calculation needs is the off-equilibrium response, and what a tolerance calculation needs is that response together with the perceived one, since the tolerance is what the public believes the central bank will accept before it acts. No design identifies the first, and none claims to; the perceived-rule designs identify the second up to a belief model. For our purposes the accumulated dispute is the datum: forty years in, the profession does not agree on whether $\varphi_\pi$ is identified, on whose information set it should be estimated, on whether its residual is a shock, or---since the lower bound---on whether the instrument it governs is even observed. And the tolerance dial sits behind all of it: the data see only the product $\varphi_\pi(1-\lambda)$, so every ambiguity above propagates into $\lambda$ with a sign chosen by the maintained assumption, while the rule literature's own frontier---perceived rules, source-dependent rules---now measures the dial's two faces without being able to point-identify either \citep{vansteenberghe2026tolerable}.

{\footnotesize
\setlength{\LTcapwidth}{\textwidth}
\begin{longtable}{>{\raggedright\arraybackslash}p{2.0cm}>{\raggedright\arraybackslash}p{4.0cm}>{\raggedright\arraybackslash}p{4.4cm}>{\raggedright\arraybackslash}p{1.8cm}>{\raggedright\arraybackslash}p{2.1cm}}
\caption{Seventy years of identification, II: the rule ($\varphi_\pi$)---what each generation identified, in which sense, and the bias the next generation named. Same reading as Table~\ref{tab:idhistory_curve}; the object column records whose rule was estimated (the central bank's or the public's), what was recovered (a coefficient, a residual, a stance, a region), whether it lies on or off the equilibrium path, and the data vintage.}\label{tab:idhistory_rule}\\
\toprule
Generation & What the generation identified & What the next generation showed & Object & Sense claimed $\to$ shown \\
\midrule
\endfirsthead
\multicolumn{5}{l}{\emph{Table~\ref{tab:idhistory_rule}, continued}}\\
\toprule
Generation & What the generation identified & What the next generation showed & Object & Sense claimed $\to$ shown \\
\midrule
\endhead
\bottomrule
\endlastfoot
Reaction functions, 1963–90 & the weights on the authorities' goals \citep{dewald1963objective} & ``neither of those conditions has been met'': not robust to ``arbitrary differences in specifications'' \citep{khoury1990federal} & the Fed's rule · coefficients · on-path · revised & S1 $\to$ not even S1 \\
\addlinespace[2pt]
Calibration, 1993–99 & $\varphi_\pi = 1.5$ ``fits \ldots{} remarkably well'', ``need not be a mechanical formula'' \citep{taylor1993discretion}; estimated by era, ``from below one to above one'' \citep{taylor1999historical} & the fit is to revised data \citep{judd1998taylor,orphanides2001monetary}; the equation ``is actually a reduced form'' with a ``very large'' serially correlated error, ``especially in the gold standard'' period \citep[n.~9]{taylor1999historical} & the Fed's rule · coefficient · on-path · revised & S1 by calibration, no claim to correct; S1 as an S4-test $\to$ S1, a reduced form fitted to revised data \\
\addlinespace[2pt]
GMM regimes, 1998–2000 & forward-looking responses \citep{clarida1998monetary}; 0.83 pre-Volcker, 2.15 after, and ``self-fulfilling fluctuations are possible'' before 1979 \citep{clarida2000monetary} & on real-time data the response exceeds one in both periods \citep{orphanides2004monetary}; the post-1979 set covers active and passive \citep{mavroeidis2010monetary}; the regression recovers $\rho$, not $\varphi$ \citep{cochrane2011determinacy} & the Fed's rule · coefficient · on-path · revised & S3-G used as S4 $\to$ S3-G conditional on the vintage; S5 after 1979; S1 \\
\addlinespace[2pt]
Determinacy estimand, 2003–26 & posterior weight on the indeterminacy region \citep{lubik2004testing}; the stakes \citep{woodford2003interest} & the region is the model's: switching \citep{davig2007generalized,farmer2010generalizing,barthelemy2019monetary}, trend inflation \citep{coibion2011monetary,hirose2020monetary}, the pricing block---with a vertical long-run curve ``no amount of output-gap \ldots{} targeting can substitute'' \citep{meyergohde2026informational}; in the determinate region $\varphi$ is absent from the likelihood unless the disturbances are restricted \citep{cochrane2011determinacy} & the Fed's rule · region · off-path · revised & S3-FI/P + S5 $\to$ S5 conditional on the model \\
\addlinespace[2pt]
Real-time data, 2001–17 & on Greenbook forecasts \citep{orphanides2001monetary}; before and after Volcker, ``a strong reaction to inflation forecasts during both periods'', the 1970s a story of perceived gaps \citep{orphanides2004monetary} & with drifting coefficients the response is weak only in the later 1970s \citep{boivin2006has}; the gap and the intercept are latent objects, the gap with persistent revisions \citep{orphanides2002unreliability}, the intercept imprecise \citep{holston2017measuring} & the perceived rule · coefficient, stance · on-path · real-time & S1/S3-G of the perceived rule; S4-test passed for inflation, failed for the gap $\to$ S1 real-time, time-varying \\
\addlinespace[2pt]
Weak identification, 2008–11 & identification-robust sets \citep{mavroeidis2010monetary} & pre-1979 passivity robust, post-1979 activeness inconclusive---``good policy removes the possibility of sunspot dynamics'' that would identify it \citep[p.~497]{mavroeidis2010monetary}; sources of instability with ``offsetting effects on output volatility'' \citep{inoue2011identifying} & the Fed's rule · coefficient · on-path & S3-G $\to$ S5 after 1979; S3 for the masked change \\
\addlinespace[2pt]
Inertia, 2002–12 & $\rho \approx 0.8$ as deliberate smoothing & ``the illusion of monetary policy inertia''---persistent shocks \citep{rudebusch2002term}; the nested interval ``8 percent to 88 percent'' \citep{english2003interpreting,rudebusch2006monetary}; ``the evidence strongly favors the interest-smoothing explanation'' \citep{coibion2012why} & the Fed's vs.\ the perceived rule · coefficient vs.\ residual · off-path check & S3-G $\to$ S5 inside the first-order rule; a point only from what that nesting leaves out \\
\addlinespace[2pt]
Residual = shock, 1998–2023 & the OLS residual is the shock under recursiveness \citep{christiano1999monetary}; the Greenbook-purged residual \citep{romer2004new}; the 30-minute surprise, with its information effect \citep{nakamura2018high} & VAR reaction functions ``implausible and misspecified'' and their residuals barely correlated with futures-market shocks \citep{rudebusch1998do}, the second charge answered in \citet[pp.~104--108]{christiano1999monetary}; the surprise mixes policy with central bank information shocks \citep{jarocinski2020deconstructing,mirandaagrippino2021transmission} or with a ``Fed response to news'' \citep{bauer2023alternative} & the Fed's vs.\ the perceived rule · residual · on-path & S2a $\to$ S2a, conditional on the decomposition maintained \\
\addlinespace[2pt]
System level, 2004–11 & the rule identified jointly with the model and priors, ``three other parameters are clearly not identified'' \citep{smets2007shocks} & ``the unidentifiability of the Taylor rule coefficients in a simple New Keynesian model pointed out in Cochrane (2007)'', the medium-scale rule identified \citep{iskrev2010local}; in a small model, one of the rule's parameters ``needs to be further restricted'' \citep{komunjer2011dynamic}; equivalence ``widespread'' \citep{canova2009back}; the prior's weight already estimated \citep{delnegro2004priors}; the size of nominal frictions not discriminated by the data \citep{delnegro2008forming} & the Fed's rule · coefficient · on-path & S3-P $\to$ S3-P, on an S5 ridge in the small models \\
\addlinespace[2pt]
Drifting and switching, 2001–11 & activism drifting \citep{cogley2001evolving,primiceri2005time}; three phases, joined by gradual drift \citep{boivin2006has}; ``time variation in disturbance variances only'' \citep{simszha2006were} & coefficient and variance regimes observationally close: ``also compatible with the `good policy' hypothesis'' \citep{benati2009var}; the coefficient priced by bond markets \citep{ang2011monetary} & the Fed's rule · coefficient vs.\ residual variance · on-path & S3-P; S3-G single-equation; S3-FI $\to$ S1 time-varying; S5 at the system level \\
\addlinespace[2pt]
Non-\hspace{0pt}identification, 2006–21 & --- & ``not identified without unrealistic assumptions''; the regression ``will estimate the disturbance serial correlation parameter $\rho$ rather than the Taylor rule parameter $\varphi$''; determinacy ``depends entirely on what the Fed would do away from the [$\pi^*$] equilibrium'' \citep{cochrane2011determinacy}, restated in \citet[pp.~364, 488]{cochrane2023fiscal}; the replies restore S3 by an added assumption---a rule form \citep{sims2008identification}, private-sector inertia \citep{carrillo2008comment}, shock restrictions \citep{backus2020identifying}, a small policy-shock variance under determinacy \citep{carvalho2021taylor} & the Fed's rule · coefficient vs.\ residual · off-path & S3-G used as S4 $\to$ S1, an S4-level negative; S3 under an added assumption \\
\addlinespace[2pt]
Lower bound, 2013–16 & the shadow rate as the stance \citep{wu2016measuring,krippner2013measuring} & ``sensitive to minor choices in their estimation \ldots{} should not be presumed \ldots{} quantitatively useful'' \citep{krippner2020note} & the Fed's stance · instrument · on-path & S1 measurement $\to$ S1, a model-constructed generated regressor \\
\addlinespace[2pt]
Perceived rules, 2011–26 & the market-perceived rule from news \citep{hamilton2011estimating}; perceptions from forecasts, ``imperfect information about the rule'' \citep{bauer2024perceptions}; misperception vs.\ information \citep{sastry2026disagreement}; bond-market views \citep{bocola2024bond} & ``the perceived policy framework may differ from the Fed's historical behavior'' and diverges under forward guidance \citep{bauer2024perceptions}; the perceived inflation coefficient rose after liftoff in 2022, learned from actions \citep{bauer2024changing}; the object is a belief, not the primitive---the tolerance dial's ground \citep{vansteenberghe2026tolerable} & the perceived rule · coefficient · the expected path & S2b; S1 in forecasts; S2a in its updating; S3 in a belief model $\to$ beliefs about the rule \\
\addlinespace[2pt]
Targeted and era fits, 2024–26 & ``nearly fourfold'' stronger response to demand- than to supply-driven inflation \citep{hofmann2026targeted}; the rule fits by era \citep{nakamura2025beyond} & the bias signed, not removed; the decomposition maintained \citep{hofmann2026targeted}; under a rule written around a target path, the Taylor principle ``applies only to off-equilibrium paths that are not observed by an econometrician'' \citep{gali2025commentary} & the Fed's rule · two coefficients · on-path & S1 by OLS on decomposed regressors; S1 fits by era $\to$ S1, the decomposition maintained; the S4 object off-path \\
\addlinespace[2pt]
ECB, 2000–25 & Taylor's coefficients fit EMU averages \citep{gerlach2000taylor}; GMM \citep{gerdesmeier2003empirical}; expectations data \citep{sauer2007using}; real time \citep{belke2011does}; a calibrated difference rule and its residual \citep{orphanides2013complexity,bletzinger2017lower}; ``two regimes'' argued from expectations \citep{rostagno2021monetary} & the sign of the verdict specification-dependent; shifts in 2008 and back in 2010 \citep{gerlach2014ecb}; the stated rule---``appropriately forceful or persistent \ldots{} large, sustained deviations \ldots{} in either direction'' \citep{ecb2025strategy}, cf.\ \citet{ecb2003strategy,ecb2021strategy}---is S4 by construction and no estimate tests its off-path part; the regime dating cited from the companion \citep{vansteenberghe2026tolerable} & the ECB's rule · coefficient, residual, stated clause · on-/off-path & S1; S3-G; S1 in expectations; S4 asserted $\to$ S3-G and S1 on-path, with shifts; the S4 clause untested \\
\addlinespace[2pt]
\end{longtable}
}

\section{Where the two programs meet}
\label{sec:joint}

Sections~\ref{sec:curve} and~\ref{sec:rule} followed the slope of the Phillips curve and the rule's response to inflation one at a time. This section reads them together. Its claim is that the separation belongs to the telling, not to the problem. Nearly every design that identifies one of the two coefficients does so by borrowing a condition from the other side, and the condition it borrows is one the other program has not established. The borrowing is regular enough to tabulate (Table~\ref{tab:designs}). The subsections take the joins in turn and end with what any quantity built from both coefficients inherits.

\subsection{The three regularities}

Read side by side, Table~\ref{tab:idhistory_curve} for the curve and Table~\ref{tab:idhistory_rule} for the rule show three regularities. The first is that the corrections relabel rather than refine. When a generation gave way, its estimate was seldom overturned as a number; what changed was the object the number was taken to measure. The slope fitted in the 1960s was reread, after \citet{friedman1968role} and \citet{phelps1967phillips}, as a slope conditional on the expectations of its period; the flat curve of the 2010s was reread by \citet{mcleay2020optimal} as the imprint of the policy that generated the data; the passive pre-Volcker response of \citet{clarida2000monetary} was reread as a property of the revised data it was fitted on: the response estimated on the real-time data available to the Committee---the method of \citet{orphanides2001monetary}---was not passive \citep{orphanides2004monetary}. In the key of Section~\ref{sec:senses} each correction is a demotion of sense: a coefficient offered or read as structural, or as invariant to policy, is shown to be a fitted relation, or a coefficient that holds only inside a maintained model, given an assumption the earlier generation had not tested. %

The second regularity is that the two programs do not fail one at a time, and the two results this survey's companion takes from them can be stated as one. \citet{mcleay2020optimal} show that when the bank follows a targeting rule, the regression of inflation on the output gap returns the slope of the rule: the result concerns the curve, but its cause lies in the bank's conduct. \citet{cochrane2011determinacy} shows that in the determinate equilibrium of the New Keynesian model, the response the Taylor principle requires is never exercised, so that a regression recovers the persistence of a disturbance instead---in the model's simplest version, of the policy disturbance (p.~573)---and that the response could be recovered only under two assumptions, the second about the model in which the curve sits and the equilibrium the bank selects (p.~595): the result concerns the rule, but what it would take to identify it lies largely on the curve's side of the model. Put together, the covariation that would identify the slope is produced by departures from the rule, and the departures from equilibrium that would identify the response are never displayed, because the curve and the rule jointly exclude them. A reading of either coefficient is a reading of both, and the two readings are not independent draws. %

The third regularity is that neither program has stopped. The designs of the last decade---regional and firm-level panels, volatility-based vector autoregressions, rules read from markets and forecasters, rules whose response depends on whether inflation comes from demand or from supply---still work in the register of point identification, most of them for a narrower object than their predecessors named: a regional coefficient, a firm's pass-through of cost, a perceived response \citep{hazell2022slope,gagliardone2025anatomy,brunnermeier2021feedbacks,bauer2024perceptions,hofmann2026targeted}. The point is not that these designs are weaker than those they replaced but that their conditioning can be read off the page, and that the side each leaves unidentified is the side the other program set out to identify. %

\subsection{Simultaneity, stated by its authors}

The difficulty is older than either program, and the earliest printed statement this survey can verify concerns stabilization multipliers, not the Phillips curve. \citeauthor{goldfeld1972some}, who date the notion to the oldest reduced-form studies (\citeyear{goldfeld1972some}, p.~593, n.~3), put it as a paradox of success: ``typically, an authority that is conducting an effective stabilization program will appear to have a very small (and statistically insignificant) multiplier in simple reduced-form experiments, and, conversely, an ineffective authority will get a large (and statistically significant) multiplier'', so that, when neither the multipliers nor the authorities' behavior is known a priori, ``the coefficients of a partial reduced-form equation hopelessly entangle the astuteness of the authorities with the relative sizes of the multipliers'' \citep[p.~588]{goldfeld1972some}. %
Their remedy was to model the authorities' reaction functions \citep[p.~593]{goldfeld1972some}. %
\citet{sargent1976observational} gave the equivalence its sharpest form: two reduced forms that ``have identical residuals and so fit equally well'' \citep[p.~635]{sargent1976observational} part company only after a change of rule, so no reduced form estimated over a sample drawn under one rule, at a given sampling interval, can choose between them (p.~637). %

The mechanism fits in two equations, worth writing down because most designs in Table~\ref{tab:designs} can be read as restrictions on them. Let inflation obey $\pi_t=\kappa x_t+u_t$, with slope $\kappa>0$ and a cost-push disturbance $u_t$, and let the output gap obey $x_t=d_t-a\pi_t$, where $d_t$ collects the disturbances to demand and to policy and $a\geq 0$ is the contemporaneous response of the demand-and-policy block to inflation. Take $d_t$ and $u_t$ to be mutually independent and serially uncorrelated, with variances $\sigma_d^2$ and $\sigma_u^2$, and leave out expectations and lags. Solving the pair and projecting inflation on the gap gives the population coefficient of the naive regression,
\begin{equation}
\frac{\operatorname{Cov}(\pi_t,x_t)}{\operatorname{Var}(x_t)}=\frac{\kappa\sigma_d^{2}-a\sigma_u^{2}}{\sigma_d^{2}+a^{2}\sigma_u^{2}}.
\label{eq:joint}
\end{equation}
With $a=0$ the regression returns $\kappa$ whatever the variance of cost-push shocks, because the gap then moves with demand alone. Raising the response from zero pulls the coefficient down from $\kappa$, and it turns negative once $a\sigma_u^2$ exceeds $\kappa\sigma_d^2$; when the block has no disturbance of its own the coefficient is $-1/a$, the slope of the policy block, whatever the value of $\kappa$. If the gap is regressed on inflation instead, the estimate of the block's response $-a$ is off by $(1+a\kappa)s_d/\kappa$, where $s_d$ is the share of the variance of inflation due to $d_t$. The derivation is ours and deliberately minimal. Its published statement is \citeauthor{mcleay2020optimal}'s, in which the block is a targeting rule that keeps inflation proportional to minus the gap---the case without a demand disturbance, where $-1/a$ is the targeting rule's slope---and the diagnosis is explicit: ``The identification problem is a simple case of simultaneity bias'' \citep[p.~208]{mcleay2020optimal}. %
Their two-shock expression, eq.~(12), adds a disturbance to the targeting rule itself and reaches the verdict of equation~\eqref{eq:joint}: ``The size of the simultaneity bias to each equation depends on the relative variances of the shocks'' \citep[p.~214]{mcleay2020optimal}. %

What makes the statement a joint one is the identity of that second disturbance. The shocks to the targeting rule, they write, ``correspond closely to the typical interpretations of identified monetary policy shocks'' \citep[p.~213]{mcleay2020optimal}, and when cost-push shocks are small ``these movements in the targeting rule now trace out the Phillips curve'' \citep[p.~215]{mcleay2020optimal}: the variation that identifies the curve is the rule's residual. %
\citet{delnegro2020whats} write the same system with the policy response on the demand side and draw the conclusion for both coefficients at once: ``When demand and supply shocks cannot be directly observed, the correlation between inflation and the output gap is not informative on either $\phi$ or $\kappa$, as in the classic identification problem'' \citep[p.~325]{delnegro2020whats}. %
\citeauthor{carvalho2021taylor} published the rule's side of the same matrix: the least-squares bias in the rule's inflation coefficient ``is proportional to the fraction of the variance of $\pi_t$ due to monetary shocks'', with the curve's slope in the denominator of their expression, so that its size ``depends also on other structural parameters of the economy'' \citep[p.~142]{carvalho2021taylor}. %
The variance share that matters is the same in both readings. A bank whose rule accounts for most of the movement in its instrument leaves, in McLeay and Tenreyro's summary of \citet{ramey2016macroeconomic}, ``only a limited amount of true exogenous variation in the data, leading to weak instrument issues'' for the curve \citep[p.~233]{mcleay2020optimal}; the same bank makes the rule's own regression bias small, which is the ground on which \citeauthor{carvalho2021taylor} defend least squares \citep[p.~141]{carvalho2021taylor}. %
One share, read twice, returns a favorable verdict to one program and an unfavorable one to the other. The verdicts do not conflict, and they cannot be separated.

\subsection{The determinacy join}

The rule's program meets the curve at the boundary that decides whether a response is large enough. \citet{clarida2000monetary} estimated a forward-looking rule by GMM with instruments dated before the decision, on the maintained condition that the rule's residual is orthogonal to them \citep[p.~153]{clarida2000monetary}, and judged the pre-1979 estimate against a model whose Phillips-curve slope was calibrated rather than estimated. A footnote marks where the curve enters: ``the unit threshold value for $\beta$ obtains exactly only when $\gamma = 0$'' \citep[p.~171, n.~33]{clarida2000monetary}, $\beta$ and $\gamma$ being the rule's responses to inflation and to the gap; the same footnote calls the deviation from unity ``almost negligible'' (p.~171, n.~33) at a slope set to $0.3$ from a literature range of $0.05$ to $1.22$ (p.~170, n.~32). Smaller values of the slope or of the discount factor ``could alter the conclusion that there was indeterminacy before Volcker'' \citep[p.~494]{mavroeidis2010monetary}. %
In the three-equation model the condition for a unique bounded equilibrium is, for non-negative responses, $\varphi_\pi+(1-\beta)\varphi_x/\kappa>1$, with $\beta$ now the discount factor \citep{woodford2001taylor}: whenever the rule responds to the gap, whether a pair of responses is active depends on the slope of the curve---``determinacy now depends upon the output response coefficient $\phi_x$, and not solely upon the inflation response coefficient $\phi_\pi$'' \citep[Proposition~4.3, pp.~254--255, where the output coefficient is annualized]{woodford2003interest}. %
\citet{lubik2004testing} built the dependence into estimation. Their boundary between the determinacy and indeterminacy regions contains $\kappa$ (their eq.~31), their posterior probability of determinacy is computed jointly with the curve's parameters, and they give the reason: ``In most models the indeterminacy region is a complicated function of several parameters, not just the parameters of a monetary policy reaction function'' \citep[p.~191]{lubik2004testing}. They also give the limit: ``In the directions of the parameter space in which there is no identification, that is, where the likelihood function is flat, the prior density is not updated'' \citep[p.~191]{lubik2004testing}. %

What matters here is what \citet{cochrane2011determinacy}, whose result Section~\ref{sec:rule} reviewed, says identification of the response would take: two assumptions. The first is that the bank's reaction to deviations from the equilibrium it has chosen equals the relation between equilibrium interest rates and equilibrium inflation and output. The second concerns the model around the rule: ``The model and Fed's choice of equilibrium \ldots\ must be such that equilibrium quantities can be expressed in the `autoregressive' representation (39), with parameters $\varphi^*$ in the zone of determinacy (explosive eigenvalues), and with the error $x_{it}$ orthogonal to something we can use as instruments'', and ``Many models and many equilibria of a given model do not have this property'' \citep[p.~595]{cochrane2011determinacy}. %
The zone of determinacy the second assumption invokes is the region whose boundary, in the three-equation model, $\kappa$ helps draw. The rule's identification is thus borrowed from the private block twice, once for the form of the equilibrium and once for the location of the boundary, and the replies confirm the borrowing: identification returns under a different assumption on the rule \citep{sims2008identification}, or when the private block carries inflation inertia or habit formation \citep{carrillo2008comment}. %

Weak identification closes the circle. \citet{mavroeidis2010monetary} found that identification-robust methods cannot pin down the post-1979 reaction function, and located the first of his three explanations (p.~502) in the expectations the curve contains: good policy makes expected inflation and output less variable, and ``After 1979, sunspot dynamics may have subsided, which explains why the reaction function is poorly identified'' \citep[p.~497]{mavroeidis2010monetary}. %
The rule's identifying variation is the variability of the curve's expectations, and a rule that works removes it. This is the weak-instrument point of the previous subsection, seen from the rule's side. %

The rule's missing variation can also be sought outside the equilibrium dynamics. \citet{hack2024identification} instrument the systematic component itself, combining a measure of the Committee's composition of hawks and doves with the mechanical rotation of voting rights; what they identify is the effect of systematic policy on the propagation of shocks rather than the response coefficient, but their instrument is the kind of exogenous variation in the rule that the determinacy join lacks. %

\subsection{The likelihood join}

For the curve, the system is not a choice: in the three-equation model closed by a Taylor rule, ``the hybrid NKPC cannot be identified by single-equation GMM'', persistent shocks and interest-rate smoothing included, and identification ``must rely on cross-equation restrictions in this system'' \citep[pp.~532--534]{nason2008identifying}. Full-information estimation puts the two coefficients in one likelihood \citep{smets2003estimated,smets2007shocks}; the diagnostics that followed, surveyed by \citet{schorfheide2013estimation} and \citet{fernandezvillaverde2016solution}, show where the data cannot share out the variation. \citet{smets2007shocks} estimate the rule and the pricing block of a medium-scale model jointly and report that ``Three other parameters are clearly not identified'', among them the curvature of the goods-market aggregator, which they fix \citep[p.~592]{smets2007shocks}; that curvature is one of the parameters that translate price stickiness into the slope of the curve. %
\citet{iskrev2010local} checked the rank of the Jacobian of the model's moments and found the pricing block's own parameters entangled---``the curvature parameters \ldots\ and the Calvo parameters \ldots\ which are locally indistinguishable in the linearized model'' \citep[p.~200]{iskrev2010local}---while on the rule's side the coefficients of that model satisfy his rank condition (p.~199), and his example of a model that cannot be identified is ``the unidentifiability of the Taylor rule coefficients in a simple New Keynesian model'' \citep[p.~193]{iskrev2010local}. %
\citet{komunjer2011dynamic} apply rank conditions, derived from the restrictions that equal spectral densities impose, to the small New Keynesian model and find both halves. On the rule, ``one of the three parameters in the Taylor rule ($\psi_1$, $\psi_2$, $\rho_r$) needs to be further restricted. This result, which was overlooked in An and Schorfheide (2007), supports the argument of Cochrane (2011) that parameters of the Taylor rule are not identified without restrictions''; on the curve, ``general statements about nonidentifiability of the Phillips curve cannot be made on the basis of rank tests evaluated at a particular $\theta_0$'' \citep[p.~2018]{komunjer2011dynamic}. %
The slope's verdict thus depends on where the model's parameters, the rule's among them, are evaluated.

\citet{canova2009back} show the same at the level of the model's solution. In their three-equation example the rule's coefficient on expected inflation and the other forward-looking coefficients ``disappear from the solution'' (p.~434), while the curve's slope can be recovered from the responses to demand and to policy shocks, and the responses to cost-push shocks ``have no information for the two parameters'', the slope and the interest sensitivity of demand \citep[ECB WP p.~12]{canova2009back}; in a larger model ``ridges are present: both responses to cost push and to monetary shocks carry little information about the correct combination of $\lambda_y$ and $\lambda_\pi$'', the rule's responses to output and inflation \citep[ECB WP p.~24]{canova2009back}. %
Their warning about what fills such gaps applies to every system estimate of the rule: ``A tightly specified prior can in fact produce a well behaved posterior distribution, even if the likelihood function has little information, giving the illusion of having collected useful evidence'' \citep[ECB WP p.~18]{canova2009back}. %

None of these diagnostics, as retrieved here, reports a ridge in the pair $(\kappa,\varphi_\pi)$ as such; what they establish is that the flat directions run through the rule's coefficients on one side and the parameters that fix the slope on the other, each verdict depending on where the other is placed. Along those directions the posterior's shape is the prior's---S3-P in the key, whatever the estimator's name. The newest system-level design makes the conditioning explicit. \citet{delnegro2020whats} separate a flat curve from an aggressive rule with an identified demand impulse and an estimated model, and call the separation ``a classic identification problem that requires economic assumptions that were not needed for the data description exercise'' \citep[p.~304]{delnegro2020whats}, their own being that the impulse shifts ``primarily the demand for labor and goods'' and does so ``to roughly the same extent before and after 1990'' (p.~327); \citet{bergholt2023did}, who give the larger weight to policy, name the assumption their own scheme rests on: ``Our identification scheme \ldots\ assumes that monetary policy is represented by a Taylor rule that is unable to mimic optimal monetary policy perfectly'' \citep[p.~24]{bergholt2023did}. The two schemes also read the rule differently: in the DSGE the response to inflation ``does not change much across samples'' \citep[p.~338]{delnegro2020whats}, while the SVAR's flatter demand curve is read as ``a more aggressive response of monetary policy to inflation movements'' \citep[p.~26]{bergholt2023did}. %
Two maintained systems return two readings of the same flattening. A third system moves both coefficients at once by adding a measurement rather than a restriction: a medium-scale model augmented with series built from the text of the Committee's transcripts estimates a weaker response of the policy rate to inflation and a flatter price Phillips curve than the same model without them \citep{casella2026structural}. %

\subsection{The instrument joins}

The newer designs replace the older systems' exclusion restrictions with other sources of identification, each carrying a statement about the other side. The first family uses volatility. \citet{rigobon2003identification} showed that a simultaneous system is identified when the relative variances of its shocks shift across regimes and its coefficients do not. \citet{rigobon2003measuring} applied the idea to the rule itself, estimating the response of policy to the stock market from periods of high and low stock-market volatility, and named the price: ``the parameters ($\alpha$, $\beta$, and $\gamma$) are stable across the covariance regimes. Note that this second assumption is implicitly made in the vast majority of related empirical studies'' \citep[p.~652]{rigobon2003measuring}. %
\citet{kajuth2012identifying} carried the method to the curve, exploiting periods in which the shocks to inflation and to unemployment change their relative size. %
\citet{brunnermeier2021feedbacks} identified a monetary-policy shock and its endogenous response from variance regimes spanning the Volcker years, and kept the 1970s in part because ``correction for time-varying variance can account for what otherwise might be interpreted as regime change in monetary policy'' \citep[CEPS WP p.~9]{brunnermeier2021feedbacks}. %
The route therefore trades an exclusion restriction for the invariance of both equations across volatility regimes---in the key, S4 assumed and tested only against models with regime-specific coefficients, the least restricted of which ``fits worse'' \citep[p.~1856]{brunnermeier2021feedbacks}, the invariance's possible failure being what the authors call ``the Achilles' heel of this approach to identification'' (p.~1849). That invariance is what the regime literature is divided on: \citet{simszha2006were} fit the data best with changes in shock variances alone; \citet{primiceri2005time}, letting coefficients and variances both drift, found a systematic reaction to inflation and unemployment that strengthened over the sample. The route's own test cannot settle the dispute, since a rejection shows only that ``at least one of the maintained assumptions---that the parameters are stable, or that the stock market or the common shock are homoskedastic---is not valid'' \citep[NBER WP 8794, sec.~4.3]{rigobon2004impact}. %

The second family instruments the curve with the rule's residual, which is what the shock literature took its object to be \citep[p.~89]{ramey2016macroeconomic}. %
\citet{romer2004new} purged changes in the intended federal funds rate of their response to the Federal Reserve's own forecasts, and \citet{barnichon2020} instrument the forward-looking curve with sequences of the resulting shocks under a condition they state in full: ``The exogeneity condition\ldots is satisfied if monetary shocks are orthogonal to (i) cost-push factors, (ii) measurement error in the output gap, and (iii) measurement error in inflation expectations'' \citep[p.~2265]{barnichon2020}. %
The first clause is a statement about the rule: its residual must carry nothing of the supply side the curve is holding fixed. The high-frequency designs replace the residual by the change in futures prices around an announcement, and \citet{gertler2015monetary} state the identifying assumption as ``news about the economy on the FOMC day does not affect the policy choice'' (p.~46). %
The information-effect literature then relaxed, one at a time, the assumption that the surprise is the rule's residual alone. \citet{nakamura2018high} and \citet{jarocinski2020deconstructing} argued that announcements move private beliefs about economic fundamentals as well as about policy; \citet{mirandaagrippino2021transmission} that a valid instrument must capture policy moves that ``are not due to the central bank's concerns about either current or anticipated changes in economic conditions'' \citep[p.~86]{mirandaagrippino2021transmission}; and \citet{bauer2023alternative} that the surprise reflects ``a difference between the Fed's actual policy response function \ldots\ and the private sector's ex ante estimate of that function'' \citep[p.~665]{bauer2023alternative}. %
Each remedy maintains a different statement about what the rule responded to---the bank's private information, the span of its forecasts, the public's reading of its response---and the curve's instrument inherits whichever statement is maintained. The same conditioning can be imposed openly: in vector autoregressions identified by sign and narrative restrictions, restrictions on the rule's coefficients sharpen the monetary shock and stabilize the estimated Phillips multiplier \citep{castelnuovo2025monetary}, the statistic of \citet{barnichon2021phillips}---the curve identified conditional on a stated property of the rule. %

The third family removes the rule from the regression instead of instrumenting around it. \citet{hazell2022slope} estimate the slope from US states on the ground---which they credit to \citet{fitzgerald2014stable} and \citet{mcleay2020optimal}---that ``central banks cannot offset regional demand shocks using a single national interest rate'' \citep[p.~1304]{hazell2022slope}, with time fixed effects absorbing the national rate and the long-run regime; \citet{fitzgerald2024regional} use cities and states because ``monetary policy reacts only to aggregate shocks'' \citep[Staff Report 614, p.~2]{fitzgerald2024regional}. %
The differencing takes the national rule out of the regional regression, but the regional coefficient it leaves is not yet the aggregate slope: ``typically, regional responses can only be translated into aggregate responses through the lens of a fully specified general equilibrium model'' \citep[p.~66]{nakamura2018identification}, and in that model the rule returns as a maintained equation---``Interest rates are set at the union level according to a monetary policy rule'' \citep[p.~1790]{beraja2019aggregate}. %
Firm-level designs take the same step one level down. \citet{gagliardone2025anatomy} leave aggregate data because of the downward bias that equation~\eqref{eq:joint} describes \citep[p.~3942]{gagliardone2025anatomy}; they identify the pass-through of marginal cost into prices with sector-by-time fixed effects, and return to the output-based slope through the elasticity of marginal cost to output, which they find low in the moderate-inflation sample they study. %
On this survey's reading, the rule's stabilization of demand, removed from the firm-level regression, re-enters through that mapping. In each case the rule is differenced out of the estimating equation and restored in the aggregation, which is where the identification of $\kappa$ is decided.

A last route is open rather than taken, and what follows is this survey's reading, not a result in the literature. Identification at infinity, one of the meanings in \citeauthor{lewbel2019identification}'s catalog, recovers a parameter from observations extreme enough for the confounding to vanish \citep[section~8.2, p.~890]{lewbel2019identification}. %
Equation~\eqref{eq:joint} has a limit of this kind: as the variance of the disturbances to the policy block grows relative to that of cost-push shocks, the naive coefficient converges to $\kappa$. The nearest published instance is the event-study limit of \citeauthor{rigobon2004impact}---``In the limit, if the variance of the monetary policy shock becomes infinitely large relative to the variances of the other shocks \ldots\ then the bias goes to zero'' \citep[p.~1557]{rigobon2004impact}---whose estimator rests on a relative increase in that variance rather than on the limit itself, the property of the OLS estimate they call, after \citet{fisher1966identification}, ``near identification'' \citep[p.~1557]{rigobon2004impact}. %
No application to the slope of the curve or to the rule's inflation response was located. The route fits this problem, since it asks for episodes in which one side's disturbances dominate, and is fragile for the same reason: in such episodes the other side's invariance across regimes is least safe to assume.

\subsection{The conquest join}

In one class of models the join is not a by-product of estimation but the premise: the rule is what the bank does given the curve it believes in. \citet{sargent1999conquest} gave the class its standard form. His government fits a Phillips curve without expectations by least squares, sets inflation each period by solving a control problem against the fitted curve, and learns from the outcomes, while the true curve is expectational; the rise and fall of American inflation becomes the history of the government's estimate of the curve. %
\citet{cogley2005conquest} replaced the single approximating model by posterior weights over three curves and stated the link directly: ``As the central bank's beliefs (i.e., the parameters of its approximating model) evolve with the accumulation of data, so too does its policy rule'' \citep[p.~531]{cogley2005conquest}; their ``empirical exploration consists of a calibration'' \citep[p.~538]{cogley2005conquest}. %
\citet{primiceri2006why} estimated the class by maximum likelihood on inflation and unemployment alone: ``policymakers estimate the natural rate and the other coefficients of the model in every period and treat these estimates as true values'' \citep[p.~875]{primiceri2006why}, and the model's policymakers face an identification problem of their own, which he assumes they solve by imposing ``the prior belief that, on average, unemployment is equal to its natural rate'' \citep[p.~878]{primiceri2006why}. %
\citet{sargent2006shocks} put the mechanism in one sentence---``At each date $t$, the monetary authority updates its beliefs about the Phillips curve and then computes a first-period action recommended by a `Phelps problem' '' (p.~1193)---and let the estimation ``reverse engineer a set of government beliefs that can explain the low-frequency swings in U.S. data'' (p.~1196). %

What the estimated members of the class identify is a sequence of perceived curves, conditional on a maintained true curve and a maintained loss, with the policy instrument latent; the rule is a function of that sequence. Within the class, the regime question of Section~\ref{sec:rule}---whether the rule's coefficients changed around 1979 or only the variances of the shocks did---becomes a question about the bank's estimation of the curve, since a date at which the rule changed is a date at which the fitted curve changed; on this reading the two verdicts of the time-series evidence, variances alone \citep{simszha2006were} and a strengthening response \citep{primiceri2005time}, become two statements about how fast the bank learned. The class identifies beliefs, not primitives. The perceived-rule designs of Section~\ref{sec:rule} are its mirror image: they read the rule as markets and forecasters perceive it \citep{hamilton2011estimating,bauer2024perceptions,bocola2024bond}, and what they identify is a belief about the rule, as the conquest models identify a belief about the curve. The join runs through beliefs in both directions, and no design in either class recovers the primitive the belief is about.

\subsection{What a composite built from both inherits}

The quantities policy analysis needs---a sacrifice ratio, the verdict on an episode, the tolerance a bank should show toward a supply shock---are functions of both coefficients. Were the two identified by independent designs, the uncertainty about such a quantity could be assembled from two separate ranges. The designs of Table~\ref{tab:designs} are not independent in that sense. The reading of $\kappa$ a design delivers is conditional on a statement about the rule, and the reading of $\varphi_\pi$ on a statement about the curve; one variance share decides both the strength of the curve's instruments and the size of the rule's regression bias; one slope draws the boundary at which the rule's response starts to matter. The pairs $(\kappa,\varphi_\pi)$ the data leave open are therefore not a rectangle formed from an interval for each coefficient but a region shaped by the maintained assumptions, and a composite inherits the shape. The joint failure is not a third failure added to the two. It is the statement that the two are one, read along two axes, so that a composite cannot be point-identified by combining the best available reading of each coefficient.

The survey's companion builds such a composite \citep{vansteenberghe2026tolerable}. Its optimal tolerance takes the slope through the intensity with which a supply shock is looked through, and the rule's response as the yardstick for any softening. It names the curve and the rule as two inherited walls standing between the data and its optimum; the band it reports is adopted from published readings, not identified. Section~\ref{sec:target} shows that the level from which the gap is measured carries a third entanglement of the same kind, Section~\ref{sec:optimum} follows the optimum's own ratchet under what is not known, and Section~\ref{sec:close} states what the register licenses. %

{\footnotesize
\setlength{\LTcapwidth}{\textwidth}
\begin{longtable}{>{\raggedright\arraybackslash}p{2.4cm}>{\raggedright\arraybackslash}p{2.5cm}>{\raggedright\arraybackslash}p{2.5cm}>{\raggedright\arraybackslash}p{4.3cm}>{\raggedright\arraybackslash}p{2.4cm}}
\caption{Seventy years of identification, III: where the two programs meet---what each design identifies on each side, the condition it borrows from the side it does not identify, and the sense claimed and shown (tags as in Section~\ref{sec:senses}; `$\to$' reads `claimed $\to$ shown'). The rows are compressed by design; the qualifications are in the text.}\label{tab:designs}\\
\toprule
Design & Curve-side object & Rule-side object & Conditioning borrowed from the other side & Sense claimed $\to$ shown \\
\midrule
\endfirsthead
\multicolumn{5}{l}{\emph{Table~\ref{tab:designs}, continued}}\\
\toprule
Design & Curve-side object & Rule-side object & Conditioning borrowed from the other side & Sense claimed $\to$ shown \\
\midrule
\endhead
\bottomrule
\endlastfoot
(1) Supply controls & $\kappa$ & none; the rule causes the bias & every trade-off shock observed; the bank on its targeting rule & S3 (S1 as Gordon presented it) $\to$ S1, a mix of two slopes \\
\addlinespace[2pt]
(2) Rule-residual instruments & the gap coefficient of the curve & the purged residual, as the instrument & the residual orthogonal to cost-push factors and to noise in the gap and in expectations & S2b $\to$ S3, if the residual is a shock \\
\addlinespace[2pt]
(3) High-frequency surprises & the impulse's effect; a candidate instrument & residual, information, misread response & policy unmoved by same-day news; the surprise orthogonal to what the rule answers & S2a $\to$ S2a, given the decomposition \\
\addlinespace[2pt]
(4) Regional and firm panels & a local slope; $\kappa$ only via a model & absorbed by fixed effects; restored in the model & one national rate for all regions; parameters common across them; a stable slack-to-cost mapping & local S2b or S3-G $\to$ S3 for $\kappa$ \\
\addlinespace[2pt]
(5) Demand impulse with a DSGE or a sign-restricted SVAR & the response to a demand impulse & the response within the system & the impulse is demand, propagating alike across regimes or signed on impact; a Taylor-rule block & S2a/S2b + S3-FI, or S2a + S3 $\to$ conditional on both \\
\addlinespace[2pt]
(6) GMM on the rule; robust sets & calibrated; its expectations instrument the rule & the responses; the regime verdict & a residual orthogonal to lagged data; expectations variable enough to instrument & S3-G + S4-test $\to$ S5 after 1979 \\
\addlinespace[2pt]
(7) Determinacy estimand & $\kappa$, which draws the boundary & the responses and their region & a boundary that contains $\kappa$; flat directions left to the prior & S3-FI/P + S5 $\to$ S3-P \\
\addlinespace[2pt]
(8) Non-identification and replies & the private block, maintained & the response as an off-path threat & an autoregressive equilibrium in the determinate zone, or a restricted rule or private block & S4 $\to$ S1; S3 under added assumptions \\
\addlinespace[2pt]
(9) Least squares on the rule & $\kappa$, inside the bias & $\varphi_\pi$ & a small policy-shock variance share in a maintained determinate model & S1 $\approx$ S3 $\to$ S3, given the model \\
\addlinespace[2pt]
(10) System likelihood and diagnostics & stickiness, given a calibrated curvature & coefficients on flat directions in the small models & the whole system known up to its parameters & S3-FI/P $\to$ S3-P \\
\addlinespace[2pt]
(11) Volatility regimes; regime tests & the trade-off (one application) & the response to asset prices; the policy shock & coefficients invariant across volatility regimes, the invariance on which the regime tests divide & S3 $\to$ S3, given S4 assumed \\
\addlinespace[2pt]
(12) Conquest and learning & the bank's perceived curve & the rule as the best response to it & a maintained loss and a true or candidate curve; a prior on the natural rate; the instrument latent & S3-FI, S3-P or S1 (calibrated) $\to$ S3-P for beliefs \\
\addlinespace[2pt]
(13) Perceived rules & --- & the rule as priced or forecast & a maintained rule form and risk-premium model; the perceived target held fixed on news days or across horizons & S1, S2b or S3 $\to$ beliefs about the rule \\
\end{longtable}
\vspace{-0.5\baselineskip}
\noindent \emph{Sources by row:} (1) \citealp{gordon1982inflation,mcleay2020optimal}; (2) \citealp{romer2004new,barnichon2020}; (3) \citealp{gertler2015monetary,nakamura2018high,jarocinski2020deconstructing,mirandaagrippino2021transmission,bauer2023alternative}; (4) \citealp{fitzgerald2024regional,hazell2022slope,beraja2019aggregate,gagliardone2025anatomy}; (5) \citealp{delnegro2020whats,bergholt2023did}; (6) \citealp{clarida2000monetary,mavroeidis2010monetary}; (7) \citealp{lubik2004testing}; (8) \citealp{cochrane2011determinacy,sims2008identification,carrillo2008comment}; (9) \citealp{carvalho2021taylor}; (10) \citealp{smets2007shocks,canova2009back,iskrev2010local,komunjer2011dynamic}; (11) \citealp{rigobon2003identification,rigobon2003measuring,rigobon2004impact,kajuth2012identifying,brunnermeier2021feedbacks,simszha2006were,primiceri2005time}; (12) \citealp{sargent1999conquest,cogley2005conquest,primiceri2006why,sargent2006shocks}; (13) \citealp{hamilton2011estimating,ang2011monetary,bauer2024perceptions,bocola2024bond}.
\par
}

\section{The third object}
\label{sec:target}

The two programs of Sections~\ref{sec:curve} and~\ref{sec:rule} share a third object, and it ran the same ratchet. The Phillips curve of Section~\ref{sec:curve} needs a level around which the gap is measured; the rule of Section~\ref{sec:rule} needs a number from which the deviation is read; and both need to know why that level moves. For most of the period the Federal Reserve did not state the number. What the literature had instead was a moment---the autocorrelation of inflation, high in the 1970s and lower after---and a sequence of homes for it. This section follows that sequence in the register of Section~\ref{sec:senses}. The section closes on what the third object leaves to the tolerance problem: a drift that is located and priced, and a source the equilibrium data do not contain. %

\subsection{From intrinsic persistence to a drifting target, 1995--2014}

The first home was the supply relation itself. \citet{fuhrer1995inflation} found that ``the Phelps-Taylor specification implies far too little inflation persistence'' (p.~127) and supplied a contracting model in which agents care about relative real wages, so that the autocorrelation is a structural feature of wage setting, identified by limited-information maximum likelihood, around a constant inflation mean, on a small system closed by reduced-form equations for the output gap and the bill rate. The reduced-form heir of that structure was the lagged-inflation term of the hybrid Phillips curve. \citet{gali1999inflation} estimated it by GMM and found backward-looking price setting ``statistically significant'' but ``not quantitatively important'' (p.~195); \citet{fuhrer2006intrinsic} read the same term the other way, as ``the dominant source of persistence'' (p.~49). %
Both identified the coefficient conditional on a constant target (S3-G, by GMM and by limited-information likelihood); the assumption on the mean was not among the things the instruments could test. %

The second home was indexation. \citet{christiano2005nominal} required firms that do not re-optimize to index to lagged inflation, which places a unit-weighted lag in the pricing equation as a structural device, and chose the parameters to match the response to an identified monetary impulse---an S2a object backed out into S3 coefficients conditional on the indexation assumption. \citet{smets2003estimated} and \citet{smets2007shocks} estimated the indexation weight by Bayesian likelihood inside a full model---a posterior mean of $0.24$ for price indexation in the 2007 model (p.~594)---with the target inert or absent: the 2003 model's persistent ``shock to the inflation objective'' ``does not seem to play much of a role'' (ECB WP 171, pp.~17, 23), and the 2007 model states that ``there is no time-varying inflation target'' (p.~586). %
In the key, S3-P conditional on an inert or absent target.

The relocation came through four doors, each of which identified a path (all but the first also a variance) and none of which tested whose drift it was. \citet{kozicki2001shifting} showed that the long-horizon forecasts a no-arbitrage term-structure model needs are heavily influenced by the ``endpoints'' of the short-rate process and that the usual assumptions ``inadequately account for historical shifts in market perceptions of the policy target for inflation'', and linked ``endpoint shifts to agent learning about shifts in long-term policy goals'' (p.~613). %
The endpoint is a latent state of a learning model, identified from the cross-equation link between the short-rate process and long yields (S3), and by the authors' own description it is the market's perception of the goal, not the goal. \citet{cogley2005drifts} let the coefficients of a VAR drift and read the drift of the local mean, of the spectrum at zero and of a policy-activism coefficient as evidence ``that monetary policy rules have changed and that the persistence of inflation itself has drifted over time'' (p.~263)---S3-P, the mean and the persistence functionals of a random-walk law and priors. %
\citet{stock2007has} found the univariate process ``well described by an unobserved component trend-cycle model with stochastic volatility'', with the standard deviation of the permanent disturbance ``approximately 0.5 percentage points at an annual rate'' before 1970, ``nearly 1.5 percentage points'' through 1983, back to its 1960s value in the mid-1980s and lower still after 1990, all for GDP inflation (pp.~3--4): what moved was the trend's innovation variance. The trend carries no owner: the paper names a change in ``the conduct of monetary policy'' as ``the obvious explanation'' of the changes but does not ``attempt to sort through these explanations'' (pp.~5--6), and ties the trend to no target and no long-run expectations. %
The fourth door is the Phillips curve itself. \citet{cogley2008trend} log-linearized the New Keynesian Phillips curve ``around a shifting steady state associated with a time-varying inflation trend'' (p.~2102) and found that ``when drift in trend inflation is taken into account, a purely forward-looking version of the model fits the data well, and there is no need for backward-looking components'' (p.~2101): the estimates of the indexation parameter ``concentrate on zero'' (p.~2102), a zero whose identification, the authors caution, ``hinges on auxiliary assumptions about features such as VAR lag length and/or the autocorrelation properties of the cost-push shock'' (p.~2114), and the autocorrelation the first two homes had housed is, as they hypothesized, ``an artifact of the interaction between drift in trend inflation and nonlinearities in the Calvo model'' that ``might be mistakenly attributed to intrinsic inflation persistence'' \citep[p.~2102]{cogley2008trend}. %
\citet{cogley2010inflation} redefined persistence on the gap to the trend and found it rising through the Great Inflation and falling after Volcker; \citet{ascari2014macroeconomics} surveyed the trend-inflation model and its consequences for the slope of Section~\ref{sec:curve}.

Read in the key, the relocation is a demotion of sense at no cost in fit: the coefficient that was S3 conditional on a constant mean goes to zero once the mean follows a random walk, and its moment is now the trend's. But the owner of the trend is not a finding. \citet{cogley2008trend} write that drift in trend inflation ``should ultimately be attributed to shifts in that target'' (p.~2102) and, in the abstract, that they ``attribute'' it ``to shifts in monetary policy'' (p.~2101); \citet{cogley2010inflation} ``associate movements in trend inflation with shifts in the Federal Reserve's target'' (p.~45); \citet{ascari2014macroeconomics} ``use the terms trend inflation, steady state inflation, and inflation target interchangeably'' (p.~680, n.~4). %
The identity trend $=$ target is the maintained assumption of the third home, stated as such by every author who uses it, and it is exactly what the next design put to a test.

\subsection{The target as a state, and whose it is, 2003--2013}

\citet{ireland2007changes} is the terminal exhibit because it is the only design that estimates the target as a state inside the full structure and then submits the attribution to a likelihood. The motivation is the pre-announcement fact: ``the Federal Reserve has never explicitly revealed the setting for its inflation target. Hence, a statistical or econometric model must be used'' (p.~1852). The target is modeled as a random walk with innovations that load on the model's two supply shocks---the bank ``systematically adjust[s] its inflation target in response to either or both of the two supply shocks'' (p.~1858)---plus an innovation of its own, whose variance is a free parameter. %
The path is identified with the help of one assumption, stated as such and motivated by Friedman's dictum: permanent changes in measured inflation cannot occur without corresponding changes in the target (p.~1859). %
Under it the smoothed target rises from ``1\textonequarter\% in 1959 to over 8\% in the mid to late 1970s before falling back below 2\textonehalf\% in 2004'' (p.~1851), %
and the backward-looking weight in price setting ``lies up against its lower bound of zero'' (p.~1864), the previous subsection's relocation reproduced inside a maximum-likelihood system. %
The point estimates then attribute the whole of the target's movement to the deliberate response and none to the exogenous innovation (pp.~1863--64), and the paper's own verdict is that the attribution does not survive the test: ``considerable uncertainty remains about the true source of movements in the Federal Reserve's inflation target, so that one cannot statistically reject the exogenous target model that depicts those movements as purely random'' \citep[p.~1864]{ireland2007changes}; the two variants are ``better-fitting and statistically indistinguishable'' \citep[p.~1866]{ireland2007changes}. %
In the key, the target's path and its innovation variance are S3-FI, identified by the cross-equation restrictions of the full model under the maintained Friedman assumption; the source cannot be identified---the likelihood leaves the deliberate and the exogenous reading open, S5 for the attribution, inside the profession's own structural model and on the postwar sample. The other designs did not run the test. \citet{schorfheide2005learning} made the target an exogenous two-state regime the public learns; \citet{delnegro2013dsge} made it an exogenous AR(1) state pinned by survey expectations. Both deliver a path; both fix the source by construction. \citet{cogley2010inflation} re-estimated the endogenous variant, without a test, and found the target's variation ``still driven primarily by the exogenous shock'' (p.~57, n.~16)---the reverse of Ireland's point estimates---before modeling the target ``as an exogenous random process'' (p.~57). %

The learning stage relocated the persistence once more, out of the target's law and into beliefs, and assigned the drift by model class. \citet{erceg2003imperfect} wrote the target as a constant plus a persistent component with an autoregressive root ``arbitrarily close to unity'' and a transitory component with a much smaller root (p.~928), let households and firms ``use optimal filtering to disentangle persistent and transitory shifts in the monetary policy rule'' (p.~915), and generated persistence and a costly disinflation from the filter rather than from a lagged-inflation term, with a Kalman gain of $0.13$ matched to survey expectations (p.~931). %
The components are exogenous; the persistence is the public's. \citet{primiceri2006why} and \citet{sargent2006shocks} put the drift in the policymaker's beliefs: policy is set ``optimally, conditional on their current beliefs'' about the natural rate, the persistence of inflation and the slope \citep[p.~867]{primiceri2006why}, and ``the monetary authority's inflation target evolves as its estimated Phillips curve moves'', so that their estimates attribute the rise and fall of inflation to ``an intricate interaction between the monetary authority's beliefs and economic shocks'' \citep[p.~1193]{sargent2006shocks}. %
Here the drift is chosen, but chosen because shocks moved the beliefs. \citet{orphanides2005decline} and \citet{orphanides2005imperfect} held the bank's target fixed and let the public learn it with a constant gain, so that a policy aimed at tight stabilization of unemployment around a misperceived natural rate ``uprooted heretofore quiescent inflation expectations'' (ECB WP p.~4) and, under an accommodating rule, ``the public's perception of the inflation target rises by 5 percentage points'' (p.~219). %
Three assignments---exogenous, the policymaker's, the public's---were each identified conditional on their belief processes, and one pair has since been ranked: on long-term survey expectations left out of the estimation, a fixed target, learned by price setters who switch to a constant gain when forecast errors persist, outperforms an inflation trend that ``evolves exogenously'' \citep[p.~29]{carvalho2023anchored}---a ranking within two maintained models, the policymaker's assignment outside it. The natural rate's own drift, the other component of the intercept and the object of the real-time literature \citep{laubach2003measuring}, is the Orphanides--Williams rung of Section~\ref{sec:optimum} and is not repeated here. %

\subsection{The deliberate reading: opportunistic disinflation, 1996--2006}

The deliberate reading has a literature of its own, and it states the moving target as a design rather than identifying it. \citet{orphanides2002opportunistic} gave the opportunistic central bank an intermediate target ``as a function of the inherited inflation rate'' (p.~53) and a loss function that ``depends on the deviation of inflation from the intermediate target, not the long-run target'' (p.~52), so that the bank ``should wait for exogenous circumstances---e.g. favourable supply shocks and unforeseen recessions---to deliver the desired additional reduction in inflation'' (p.~48). %
\citet{bomfim2000opportunistic} wrote the interim target as the minimum of its own past value and inherited inflation, reset to realized inflation whenever a favorable shock or a recession delivers progress (eqs.~2--3, pp.~709--710), and indexed the interim target's credibility by the weight the public places on it against last period's inflation (eq.~5, p.~711). %
None of this is an estimate: the interim target is the rule's definition, and its sense is that of a policy class. \citet{gurkaynak2005sensitivity} noted the ``similarities to `opportunistic disinflation'\,'' of a target that adjusts toward trailing inflation (p.~432, n.~11), and \citet{ireland2007changes} made the opportunistic response the estimated loading of the target on the supply shocks, chiefly the cost-push shock, found the point estimate on it, and reported that the model ``turns out to be statistically indistinguishable from the alternative, exogenous target variant that depicts movements in the inflation target as purely random'' (p.~1876). %
The deliberate reading is the design's stipulation and the estimate's point; it is not the test's verdict.

\subsection{The perceived target, 2005--2025}

The market-reading generation moved the object from the bank's target to the market's estimate of it and identified different things: the pricing of the drift, the anchoring moment, and---after the number was stated---the credibility of the number as a gain. It is the target-side counterpart of the perceived-rule generation of Section~\ref{sec:rule}, and its authors are as explicit that the object is a belief.

\citet{gurkaynak2005sensitivity} found far-forward rates responding to macroeconomic and policy surprises, ``consistent with a model that we present in which private agents' views of long-run inflation are not strongly anchored'' (p.~425), and fitted the response with a model in which the target ``is not directly observed by the private sector, and thus must be inferred by agents on the basis of the central bank's actions'' (p.~432)---the response an S2a object, the mechanism a calibrated learning model in which the target adjusts toward past inflation with a gain set to match the data (S1, by calibration). %
The drift in their model is the bank's own---its target ``really does vary over time''---and on its motive they are explicit: ``We do not take a stand on why the central bank might behave in this way'' (p.~432). %
\citet{gurkaynak2010does} turned the response into a criterion---if far-forward inflation compensation ``is relatively stable and insensitive to incoming economic news'', that ``would suggest'' that market participants ``have fairly stable views regarding the distribution of long-term inflation outcomes'' (p.~1210)---and compared the United States with the United Kingdom after 1997 and Sweden: the Fed's ``informal approach to a long-run inflation objective does not seem to have anchored the private sector's long-term inflation views to the same extent'' (p.~1236). %
\citet{beechey2011are} ran the same comparison across the Atlantic and named the confound of the whole moment: a systematic reaction of inflation compensation to news ``most likely owes to expected inflation or the inflation risk premium, or both'' (p.~115). %
These are S1 objects compared across regimes---an S4-test in spirit on the anchoring moment, the regime confounded by whatever else differs across countries and dates---and they concern the distribution of long-run outcomes, not the level of the target.

The level came from elsewhere. \citet{hamilton2011estimating}, estimating the market-perceived rule of Section~\ref{sec:rule}, wrote that ``the Fed's inflation target is unobservable'' and built the method ``to difference out these slowly moving latent variables'' (p.~2)---the target a nuisance, its latency named. %
\citet{bauer2024perceptions} make a related concession: the perceived target and natural rate are ``assumed to follow martingales'', and the intercept they enter ``differs across forecasters but not across horizons'' (p.~2238). %
\citet{bauer2020interest} priced the trend in yields---the shifting endpoint the sum of a real-rate trend and an inflation trend that ``can be viewed as the perceived inflation target of the central bank'' (p.~1322)---and took the level from a proxy that adds a real-time estimate of the real-rate trend to survey-based trend inflation, or estimated it from yields alone under a tight smoothness prior (pp.~1334--1335), since ``conditional on $P_t$, yields do not load on $i^*_t$'' (p.~1333): S3-FI for the pricing, the level observed or, from yields alone, identified only once the prior is imposed (S3-P). %

Once the number was stated, the latent object became its credibility, and the designs identified a gain. \citet{reis2021losing} distinguished an anchor that is ``adrift'', when expectations ``persistently change'', from one that is ``lost'', when ``they differ from the central bank's target'' (p.~307), and measured a pre-announcement drift of ``2.3 percentage points between 1967 and 1971'' by combining the cross-section of household expectations with professional and market expectations (p.~310); the level is not measured, since ``the initial level of expected inflation is not identified'' (p.~336, n.~18). %
\citet{ehrmann2021point} compared point targets, tolerance bands and ranges across twenty economies and found that while ``target ranges and tolerance bands perform better in several tests, none of the target types consistently outperforms the others'' (p.~2). %
\citet{bundick2023did} modeled long-run expectations as drifting toward realized inflation at a rate $\delta_\pi$, calibrated it at $0.27$ from the response of inflation compensation to inflation news over 1999--2011 (pp.~1314, 1318), and found the response falling after the Summary of Economic Projections and the 2012 statement, with the level of the objective an input calibrated to the pre-2012 Phillips-curve regressions (p.~1314) and not identified by the event studies (p.~1318, n.~9); there is ``no evidence of anchoring in Japan despite the adoption of a numerical inflation target'' (p.~1310). %
\citet{jorgensen2024anchored} estimated the gain of a Kalman-filter expectation inside the Phillips curve, read it as ``measuring the degree of anchoring'' (p.~3), and found it declining over 1960--2019, from about $0.3$ in 1960--1983 to about $0.1$ in 1984--2007 and to a value not statistically different from zero in 2007--2019 (pp.~7--8): most of the fall predates the 2012 statement, and the subsamples place no break at it. %
\citet{malmendier2026seemingly} then showed the moment itself to be consistent with a model in which individuals learn only from the inflation they have lived through, without anchoring their expectations to an announced target, so that the moment does not identify the target as the anchor. %
\citet{nagel2026leaning} reads a large share of the secular variation in real interest rates as policy leaning against such experience-based expectations of long-run inflation. %
A low gain says that the perceived target moves little with surprises; it does not say that the perceived target equals the stated one, and no design in this generation reports the post-2012 perceived level against the announced number with an error band. What the generation identifies, in every case, is the pricing of a drift and the sensitivity of a belief; whose drift it prices is the question the previous subsections left, renamed latency and stepped around.

\subsection{The stated target, 2003--2026}

The announcement changed the object without changing the identification problem. In January 2012 the Federal Open Market Committee stated that ``inflation at the rate of 2 percent, as measured by the annual change in the price index for personal consumption expenditures, is most consistent over the longer run with the Federal Reserve's statutory mandate'' and asserted the effect the previous generation had been measuring as an outcome: ``communicating this inflation goal clearly to the public helps keep longer-term inflation expectations firmly anchored'' \citep{fomc2012statement}. In August 2020 the tolerance around the number was stated qualitatively---after periods of inflation persistently below 2 percent, policy ``will likely aim to achieve inflation moderately above 2 percent for some time'' so that inflation ``averages 2 percent over time'' \citep{fomc2020statement}---and in August 2025 that language was withdrawn while ``the Committee's two percent longer-run inflation goal was not a focus of the review'' \citep{fomc2025statement}; the text reaffirmed in January 2026 keeps the number \citep{fomc2026statement}. %
On the euro-area side the Governing Council confirmed in 2003 its 1998 definition of price stability as inflation ``below 2\%'' while aiming ``close to 2\% over the medium term'' \citep{ecb2003strategy}, replaced the definition in 2021 by a symmetric two percent target under which negative and positive deviations are ``equally undesirable'' \citep{ecb2021strategy}, and in 2025 recognized ``the importance of appropriately forceful or persistent monetary policy action in response to large, sustained deviations of inflation from the target in either direction'' \citep{ecb2025strategy}---the stated rule whose estimated counterparts Section~\ref{sec:rule} reviews. %

In the key, these are S4 clauses asserted by the institution: a policy-invariant primitive declared rather than estimated, together with an off-path commitment---the make-up rule of 2020, the ``large, sustained deviations'' clause of 2025---whose parameters were never given a number. No estimate tests them, for reasons the designs themselves supply. The designs that estimate anything after 2012 hold the perceived target at the announced number or difference it out, so the level is an input; the designs that estimate a latent long-run object identify its gain, and a gain is not a level; and the tolerance stated in 2020 was withdrawn in 2025 without any design having identified what the public took ``moderately above 2 percent for some time'' to mean. The stated target and the stated rule are the same kind of object seen from two sides, and the survey's verdict on them is the same: the clause is invariant by construction and untested by any estimate on the equilibrium path.

\subsection{What the third object inherits}

The third object ran the ratchet of Sections~\ref{sec:curve} and~\ref{sec:rule} on a single moment and ended in the same place. The persistence of inflation was identified as a contract structure, then as an indexation weight, then as the innovation variance of a drifting trend, each time conditional on an assumption about the mean that the next generation relaxed and named: intrinsic persistence, imposed indexation, misattributed autocorrelation. The relocation to the trend is the robust result---no modern estimate needs a backward-looking term once the mean drifts, and the drift is priced in long rates and in the anchoring moment---and it is a demotion of sense, not a refutation, because the trend inherited the moment on the same terms the coefficient had held it. The owner of the trend was then attached by identity (trend $=$ target), by model class (exogenous components, the policymaker's beliefs, the public's perception), by design (the interim target) or by survey (the perceived target). The one design that put two owners in one likelihood reported that the data do not choose between them, while a comparison across two models ranked the public's perception above an exogenous trend \citep{carvalho2023anchored}. The announcement era stated the number and left its credibility as the latent object, identified as a gain whose fall the designs date differently. %

What the tolerance problem inherits is both halves. The companion paper's implicit-target equation carries the drift in its residual, where the third object's literature located it. The attribution the residual cannot make---deliberate accommodation of a supply shock or a drifting target---is the rule-side wall that paper names \citep{vansteenberghe2026tolerable}; the joint failure of Section~\ref{sec:joint} is the same entanglement seen from the curve's side, since the level the gap is measured from is the level whose motion this section could not attribute. The third object therefore adds nothing to the two programs' catalog of successes and one item to their catalog of biases: a moment successfully relocated, and a source that seventy years of equilibrium data, a full structural model and a stated number have left unidentified. %

{\footnotesize
\setlength{\LTcapwidth}{\textwidth}
\begin{longtable}{>{\raggedright\arraybackslash}p{2.0cm}>{\raggedright\arraybackslash}p{3.9cm}>{\raggedright\arraybackslash}p{4.0cm}>{\raggedright\arraybackslash}p{1.9cm}>{\raggedright\arraybackslash}p{2.0cm}}
\caption{Seventy years of identification, IV: the target---what each generation identified, in which sense, the bias the next generation named, and whose drift each design attributes by assumption. Sense tags as in Section~\ref{sec:senses} (S1 a fitted relation · S2a an identified impulse · S3 a model-conditional coefficient or latent state, by GMM or another limited-information method (S3-G), full-information likelihood (S3-FI) or priors (S3-P) · S4 a policy-invariant primitive · S5 a set); `$\to$' reads `claimed $\to$ shown'. The table is deliberately coarse; the text gives the qualifications and the page-pinned quotations.}\label{tab:target}\\
\toprule
Generation & What it identified & What the next generation showed & Whose drift & Sense claimed $\to$ shown \\
\midrule
\endfirsthead
\multicolumn{5}{l}{\emph{Table~\ref{tab:target}, continued}}\\
\toprule
Generation & What it identified & What the next generation showed & Whose drift & Sense claimed $\to$ shown \\
\midrule
\endhead
\bottomrule
\endlastfoot
Intrinsic persistence, 1995--2006 & persistence as a contract structure: ``far too little inflation persistence'' in the Phelps--Taylor model \citep{fuhrer1995inflation}; the lagged term ``not quantitatively important'' \citep{gali1999inflation} or ``the dominant source'' \citep{fuhrer2006intrinsic} & once the mean drifts, the autocorrelation might be ``mistakenly attributed to intrinsic inflation persistence'' \citep{cogley2008trend}; the backward weight ``up against its lower bound of zero'' \citep{ireland2007changes} & none (constant mean) & S3-G $\to$ S3 conditional on a constant mean \\
\addlinespace[2pt]
Indexation, 2001--07 & full indexation matched to an identified impulse \citep{christiano2005nominal}; a partial weight estimated with the objective inert, then absent \citep{smets2003estimated,smets2007shocks} & the estimates ``concentrate on zero'' under a drifting trend \citep{cogley2008trend} & none; exogenous and negligible (2003) & S2a $\to$ S3; S3-P $\to$ S3 conditional on an inert target \\
\addlinespace[2pt]
Shifting endpoints, 2001--05 & the endpoint of the short-rate process from long yields and a learning rule, tracking ``market perceptions of the policy target'' \citep{kozicki2001shifting} & the endpoint moves daily with news \citep{gurkaynak2005sensitivity}; yields price the market's estimate & perceived & S3 $\to$ S3 (the market's endpoint) \\
\addlinespace[2pt]
Drifting trends, 2005--10 & a drifting VAR mean \citep{cogley2005drifts}; a trend whose innovation variance moves \citep{stock2007has}; gap persistence \citep{cogley2010inflation} & the movement is in the permanent component's variance; the owner attached later, trend to ``shifts in the Federal Reserve's target'' \citep{cogley2010inflation} & none stated; the bank's, by assumption (Cogley--Primiceri--Sargent) & S3-P $\to$ S3; the owner not tested \\
\addlinespace[2pt]
Trend-inflation NKPC, 2008--14 & the curve around ``a shifting steady state'': the forward-looking version fits with ``no need for backward-looking components'' \citep{cogley2008trend,ascari2014macroeconomics} & trend $=$ target ``should ultimately be attributed'', used ``interchangeably''---an identity, not a finding & the bank's, by assumption & S3-G $\to$ S3 conditional on trend $=$ target \\
\addlinespace[2pt]
The target as a state, 2003--13 & a random-walk target with supply loadings and an exogenous innovation, estimated in the full structure; the path 1959--2004 \citep{ireland2007changes}; an exogenous regime or AR(1) \citep{schorfheide2005learning,delnegro2013dsge} & the deliberate and the exogenous reading ``statistically indistinguishable'' \citep{ireland2007changes}; the other designs impose the exogenous one & \hspace{0pt}indeterminate & S3-FI/P for the path $\to$ S5 for the source \\
\addlinespace[2pt]
Learning, 2003--06 & the public's signal extraction \citep{erceg2003imperfect}; the policymaker's beliefs \citep{primiceri2006why,sargent2006shocks}; the perceived target under a constant gain fitted to survey forecasts \citep{orphanides2005decline} or varied over a range \citep{orphanides2005imperfect} & each assignment fixed by the model class; the public's perception ranked above an exogenous trend \citep{carvalho2023anchored} & exogenous / deliberate given beliefs / perceived & S3 $\to$ S3 conditional on the belief process \\
\addlinespace[2pt]
The deliberate reading, 1996--2006 & the interim target as a function of inherited inflation, by design \citep{orphanides2002opportunistic,bomfim2000opportunistic} & estimated as the supply loading, it is the point estimate, not the verdict \citep{ireland2007changes} & deliberate, by design & a policy class $\to$ S3 point; S5 verdict \\
\addlinespace[2pt]
The perceived target, 2005--25 & far-forward rates on news, a target ``inferred by agents'' \citep{gurkaynak2005sensitivity}; anchoring across regimes \citep{gurkaynak2010does,beechey2011are}; the target differenced out or held constant across horizons \citep{hamilton2011estimating,bauer2024perceptions}; the trend priced, its level from surveys or from yields under a smoothness prior \citep{bauer2020interest}; the anchor ``adrift'' or ``lost'' \citep{reis2021losing}; target types \citep{ehrmann2021point}; credibility as a gain \citep{bundick2023did,jorgensen2024anchored} & the moment concerns the distribution, not the level, with a risk-premium confound; the level an input or a survey; the gain's fall dated to the announcement or to the low-variance era; ``no evidence of anchoring in Japan'' \citep{bundick2023did}; the moment reproduced by learning from experience without a target \citep{malmendier2026seemingly} & perceived; the bank's motive ``not take[n] a stand on'' & S2a / S1 / S3 $\to$ S1 for the moment; S3 for the gain; the level not identified \\
\addlinespace[2pt]
The stated target, 2003--26 & the number stated, its anchoring asserted \citep{fomc2012statement}; the tolerance stated, then withdrawn \citep{fomc2020statement,fomc2025statement,fomc2026statement}; the ECB's definition, symmetric point and deviations clause \citep{ecb2003strategy,ecb2021strategy,ecb2025strategy} & no estimate tests the clause's off-path part; the perceived tolerance never identified; the number alone not the identifying event & stated & S4 clause asserted $\to$ untested \\
\end{longtable}
}

\section{The optimum under what is not known}
\label{sec:optimum}

The estimation programs of Sections~\ref{sec:curve} and~\ref{sec:rule} asked what the data can identify. A quantitative tradition running beside them since the 1950s asked how far the optimal policy rule moves when something is not known, and answered in rungs. Each rung computed the optimum conditional on something it held fixed---the parameters, the model, a family of models, the formation of expectations, the freedom of the policy rate to fall---and the next rung relaxed that conditioning and reported what changed: the ratchet of the estimation programs, run on a claim of optimality. In the key of Section~\ref{sec:senses} the claim has three grades. A \emph{point} is an optimum computed within a maintained model whose parameters and loss are treated as known, an S3 object conditional on that model. A \emph{band} is a class of rules whose loss stays near the attainable minimum in every member of a named family of models, or in every model near a reference model; the optimum cannot be point-identified across the family, and the band is what the family leaves open. A \emph{sign} is the direction of the response, or of the adjustment to it, without a magnitude. Table~\ref{tab:rungs} lists the rungs, what each held fixed and which grade survived; it is the optimality counterpart of Table~\ref{tab:idhistory_rule}. The survey states no optimum of its own.

\subsection{The ladder}

The ground floor is a theorem about what uncertainty does not do. With a linear model, a quadratic loss and additive disturbances, the optimal decision under uncertainty is the decision under certainty with the unknowns replaced by their expected values \citep{simon1956dynamic,theil1957note}. %
\citet{svensson1997inflation} derived his inflation-forecast rule as ``a straightforward application of standard certainty-equivalence in linear-quadratic models'' (p.~1118), %
and \citet[p.~608]{woodford2003interest} added the qualification the rest of the ladder turns on: ``one cannot naively apply this principle to any rule that describes optimal policy in the case of full information.'' %
Within its scope the theorem moves the uncertainty from the rule to the loss. In an estimated euro-area model, noise in the measured state leaves the unrestricted optimal rule unchanged except that it acts on estimates, while the loss rises ``from 1.03 to 1.60'' \citep[p.~101]{peersman1999taylor}. %
That form is a point, conditional on the model.

The theorem fails in three places, and the rungs above it visit each. The first is uncertainty that multiplies the instrument's effect: \citet{brainard1967uncertainty} showed that when the effect of the instrument on the target is uncertain and independent of the disturbance, the optimal response is smaller than the certainty-equivalent one. This attenuation, the origin of the ladder, is the result practice carried forward \citep{blinder1998central}, but its scope was narrow. In the reading of \citet[p.~104]{peersman1999taylor}, ``only in the special case where none of the parameters is correlated can one unambiguously show that higher uncertainty about the transmission mechanism will result in a more cautious response of the central bank to the economy's state variables'', %
and with several uncertain parameters, as \citeauthor{rudebusch2001timid} reads \citet[ch.~10]{chow1975analysis}, ``almost nothing can be said even qualitatively about how the optimal rule under model uncertainty changes relative to the optimal rule under certainty'' \citep[p.~207]{rudebusch2001timid}. %
The other two places are the ``two exceptions to the principle of certainty equivalence that operate here'' in \citet[p.~213]{rudebusch2001timid}: a simple rule, which the theorem no longer protects, in place of the fully optimal one, and a rule that responds to noisy data rather than to efficient estimates of the state \citep{swanson2004signal}. %
With no qualitative theorem for these cases, the tradition went quantitative.

\subsection{Parameter uncertainty within a model: the empty rung and its exception}

The first quantitative rung fed a maintained model's estimation covariance into the optimal-rule calculation, and three exercises found almost nothing. \citet{estrella1999rethinking}, with orthogonal uncertainty about every coefficient of a small model, reported that ``The Brainard-type adjustment---a 2.5 percent reduction---is particularly small'' (p.~422). %
\citet[p.~104]{peersman1999taylor} found that ``conventional parameter uncertainty does not seem to matter very much for the efficient instrument rules'', while noting that they had not tried euro-area models very different from their own (p.~106). %
\citet{rudebusch2001timid} found its effect on the optimal Taylor-rule coefficients ``negligible'' when only the interest-rate sensitivity is uncertain (p.~208) and ``quite modest'' overall (p.~209), and concluded that ``parameter uncertainty---even about the value of the equilibrium real rate---does not matter'' (p.~215). %
He also stated what it rests on: ``This point is fundamental to my analysis, which is predicated on the assumption that policymakers acted in an optimal manner. If I find an optimal policy rule for a particular model that matches the historical policy rule, this result is surely undercut if policymakers believed that they were optimizing in a completely different model'' \citep[p.~205]{rudebusch2001timid}. %
In the key the rung delivers an S3 point that its author demoted himself: the optimum is identified conditional on the model being the policymakers' own, and the rung's emptiness belongs to that conditioning.

The rung is empty unless model and rule are both unrestricted. In an unrestricted vector autoregression with an unrestricted rule, the policy calculated under parameter uncertainty ``can account for a considerable portion of the gradualism observed in funds rate movements'' \citep[FEDS WP abstract]{sack2000does}; %
\citet[p.~209]{rudebusch2001timid} traced the difference to the design, since an optimized unrestricted rule ``tends to have coefficients that are overly attuned to the random idiosyncracies in the estimated coefficients of a VAR''. %
\citet{soderstrom2002monetary} kept a small model, changed which parameter is uncertain, and reversed the sign of the adjustment: with uncertain inflation persistence ``it may be optimal for the central bank to respond to shocks more aggressively in order to reduce uncertainty about the future development of inflation'' (p.~125), while ``Uncertainty about the impact effect of policy still leads to more cautious policy, in accordance with Brainard's original analysis'' (p.~140). %
These are not rival estimates of one quantity; each is the adjustment belonging to one conditioning.

Noise in the measured state is where the rung yields a direction rather than a size. With a two-argument rule and a mismeasured output gap, the weight on output falls from 1.58 to 1.41 and the weight on inflation rises from 1.53 to 1.65, and ``It becomes optimal not to respond to the output gap when its standard deviation is larger than 1.15\%'' \citep[pp.~101--102]{peersman1999taylor}. %
\citet{swanson2004signal} showed the same direction for an unrestricted, fully optimal rule: once the rule is a reduced form that merges signal extraction with the response, certainty equivalence fails, and the optimal rule reacts less to a variable whose uncertainty has risen and more to the others. %
The magnitudes belong to each model; the direction---less weight on the noisy argument---recurs across the exercises and is the first sign the ladder hands upward, while the weight on the well-measured argument rises in \citet[p.~101]{peersman1999taylor} and falls in \citet[p.~213]{rudebusch2001timid}. %

\subsection{Model uncertainty within a family: the robust rule}

The next rung inverted the conditioning. \citet{levin1999robustness} treated each model's estimated parameters ``as known with certainty'' (p.~263, n.~1) and put the uncertainty on which of four models of the US economy, all built on ``the assumptions of rational expectations, short-run nominal inertia, and long-run monetary neutrality'' (p.~264), is the right one. %
The optimum traveled less well: ``rules with a larger number of free parameters can be fine-tuned to the dynamics of a specific model but often perform poorly in other models compared with the best simple rules'' (p.~264). A class traveled well: rules that set the change in the policy rate rather than its level stayed close to every member's policy frontier (p.~264). The conclusion states its own scope---simple rules taken from one model do well in the other three and are ``robust to model uncertainty \emph{within this class of models}'' \citep[p.~294, emphasis added]{levin1999robustness}. %
A point per model became a band across the family.

The printed comment tested the clause one family away. In a limited-participation model with flexible prices, whose rigidity is a financial-market friction (p.~301), \citet{christiano1999comment} found ``large portions of the parameter space in which application of an interest rate rule implies equilibrium indeterminacy or explosiveness'' and, under technology shocks, that ``a simple k percent monetary policy rule dominates'' Rotemberg and Woodford's preferred rule (p.~316). %
The danger was least for an inertial rule responding strongly to inflation and weakly to output (p.~301): the band did not survive the move; what remained was a corner of least risk, and with it the sign of the inflation response. %

The family was then widened. \citet{levin2003performance} added a small micro-founded model to four estimated ones: the robust rules respond to the inflation forecast about a year ahead and to the current output gap with substantial inertia, while longer-horizon rules are prone to indeterminacy. %
\citet{levin2003robust} took three reference models that disagree on expectations formation and built-in persistence: ``simple rules incorporating a moderate degree of interest rate smoothing yield remarkably robust performance in all three reference models as long as the loss function places nontrivial weight on stabilizing both output and inflation'', but ``under strict inflation targeting, there is no simple rule that yields robust performance across all three models'' (p.~948). %
The band is conditional on the loss as well as on the family.

How the members are weighed is a further conditioning. \citet{brock2003policy} held that ``conditioning policy evaluation on a particular model ignores the role of model uncertainty in the overall uncertainty that surrounds the effects of a given policy choice'' (p.~236) and averaged across models. %
The minimax alternative insures against the worst member and, as \citet[p.~849]{taylor2010simple} note, ``can be very sensitive to outlier models''. %
Their survey gives the tradition's verdict on the rung---``the robust policy differs significantly from each optimal policy by having a modest response to the output gap that is suboptimal in each model, but highly costly in none'' (p.~848)---and its reach: ``some robustness exercises yield conclusions that are contradicted by an otherwise similar exercise using a different set of models'' (p.~849). %

\subsection{Doubt without a named family}

Robust control dropped the list of models and kept its center. In \citeauthor{hansen2001robust}'s construction, ``Robust-control theory specifies Q by taking a single `approximating model' and statistically perturbing it'' \citep[p.~60]{hansen2001robust}, $Q$ being the models the decision maker entertains; the perturbations ``allow its shocks to feed back on state variables arbitrarily'', and one penalty parameter indexes models that the available sample cannot easily tell apart from the approximating one (p.~65). %
The rule is chosen to do best against the least favorable of those models. \citet{hansen2008robustness} develop it at book length.
This is the formal ceiling of the ladder, its most general statement of doubt, and it still has a center: the result is a band around a point, carrying whatever sign the reference model gives it. \citet{levin2003robust} drew the consequence: most studies had followed Brainard ``in focusing on the case in which the policymaker has a single reference model and the true economy lies within a specified neighborhood of this model'' (p.~946), and ``the robustness of such rules may be somewhat illusory, because policymakers actually face a much greater degree of model uncertainty'' (p.~969). %

In applications, the direction of the robust adjustment depended on how the doubt was written: \citet{onatski2003modeling} suggested that the aggressiveness of robust rules was likely to stem from formulations that overweight uncertainty about low-frequency dynamics. %
Brainard's attenuation and robust control's aggressiveness are adjustments of opposite sign to one certainty-equivalent response, each conditional on what it holds fixed; the response keeps its sign under both.

\subsection{Misperceived natural rates and learning}

The next rung relaxed two things held fixed so far: knowledge of the natural rates and a given scheme of expectations formation. \citet{orphanides2002robust} made policymakers uncertain about the processes generating the natural rates of unemployment and interest and about the extent of their mismeasurement, so that ``standard applications of certainty equivalence based on the classic linear-quadratic-Gaussian control problem do not apply'' (pp.~65--66). For a rule of the Taylor form, as mismeasurement grows, ``the optimal response to the unemployment gap drops (in absolute value) and approaches zero'', while ``the optimal responsiveness to inflation increases with uncertainty in this case'' (p.~92). %
The surviving rules are difference rules, which need no natural-rate estimate (Section~\ref{sec:close} quotes their definition), and the authors named what held the result in place: ``A key feature of the baseline and new synthesis models is the assumption of rational expectations'' (p.~104). %

The sequels relaxed it. When, in addition, private agents re-estimate their forecasting models each period with a constant gain, efficient policies ``call for greater policy inertia, a more aggressive response to inflation, and a smaller response to the perceived unemployment gap than would be optimal if everyone had perfect knowledge of the economy'' \citep[p.~1406]{orphanides2007robust}---the mechanism behind the reading of the 1970s in \citet{orphanides2005decline} and the learning stage of Section~\ref{sec:target}. %
Placed in an estimated model whose agents learn, the optimal-control policy computed under rational expectations fails in a stated direction: ``by implicitly assuming that inflation expectations are always well anchored, the optimal control policy responds insufficiently strongly to movements in inflation'' \citep[p.~S81]{orphanides2008learning}, while two simple rules chosen under rational expectations perform very well (p.~S81). %
In the key this is the counterpart of an S4-test, run across modeled expectations regimes rather than observed ones: the direction passes it---more weight on inflation throughout, and, where the natural rates are misperceived, less on the perceived gap and more inertia---and the magnitude, which moves with the learning gain and the variance of the natural rates, does not. The authors close the rung with the ladder's own sentence: ``robustness of any policy cannot be definitively `proved' because the policy may perform poorly in an alternative model'' (p.~S95). %

Learning also made room for experimentation. \citet{wieland2000monetary} set against Brainard's gradualism a policymaker learning the two parameters of a linear relation: a policy that separates learning from control can leave a persistent inflationary bias, whereas the optimal learner ``experiments with a tighter monetary policy'' as soon as inflation rises (p.~219). %
Later work found the motive small: a robust policymaker still has an incentive to experiment, and ``the degree of experimentation is tempered by concerns that the decision problem is misspecified'' \citep[p.~1601]{cogley2008robustness}, though distrust of the submodels alone ``increases the motive to experiment'' and, with distrust of the prior added, the robust rule has ``little or no experimentation'' (p.~1619). %
In the general Markov jump-linear-quadratic framework of \citet{svensson2008optimal} the benefits of learning ``may be substantial'' while those of experimentation are ``modest or even insignificant'' (pp.~291--292), the experimentation component often making policy slightly more aggressive and once less so (p.~276). %
Even within one framework, the sign of this adjustment is not fixed.

\subsection{The lower bound}

So far the rate could go wherever the rule sent it. The lower bound is a nonlinearity the linear-quadratic frame excludes, and the rung that admitted it answered conditional on a distribution of shocks. In the FRB/US model, \citet{reifschneider2000three} found that, outside rare severe contractions, under a Taylor rule aimed at strict price stability, the bound raises the variability of output but not of inflation. %
With shocks the size of those of the United States in the 1980s and 1990s, \citet{coenen2004price} found the bound's consequences negligible at moderate inflation targets and rising nonlinearly as the target falls. %
The next rung relaxed the shocks: after 2008, \citet{williams2009heeding} found that, under the standard Taylor rule, a 2 percent target's buffer might prove too thin if adverse conditions persisted. %
The ECB's 2021 strategy stated that, near the bound, its symmetric target requires a more forceful or more persistent response \citep{ecb2021strategy}. What survives is an ordering: the bound costs more the lower the target and the larger the shocks, by amounts the assumed shocks decide. %

\subsection{The current rung}

The latest rung took apart the one-for-one response. \citet{nakamura2025beyond} argue that the concern academic work attaches to the Taylor principle, self-fulfilling fluctuations, ``can be addressed with a shock-contingent commitment'' and is ``fragile to deviations from fully rational expectations'' (abstract). Under such a ``minimalist Taylor principle'' (p.~4) a bank may respond to demand and supply shocks as it finds optimal, and the authors give ``three reasons why a central bank with firmly anchored long-run inflation expectations optimally varies the nominal interest rate less than one-for-one with inflation in certain circumstances'' (p.~5), while keeping simple rules as ``valuable guideposts'' (p.~5). %
The principle's determinacy role concerns what the bank would do on paths the economy does not take; that is why Section~\ref{sec:rule} found the coefficient it constrains beyond the reach of equilibrium data \citep{cochrane2011determinacy}, and why an argument about optimal policy can move the principle into an off-path commitment and leave the on-path response free. %

\citet{gali2025commentary} restates the point from inside the standard New Keynesian model, where the Taylor principle is the condition for a locally unique equilibrium (p.~6). In the examples of \citet{nakamura2025beyond} that he reports, under optimal policy the coefficient on current inflation ``may very well be less than one for a plausible range of parameter settings'', though the same relation written on expected inflation has a coefficient ``unambiguously larger than unity''; in some calibrations the estimated output coefficient ``even switches sign and turns negative'' (p.~10), and the inflation coefficient itself ``could even be negative for some parameter values'' (p.~11). %
What Gal\'{\i} keeps is a sign on another object: ``a positive relationship between the long real rate and equilibrium inflation can be shown to be a robust feature of optimal policy in the New Keynesian model. I refer to that property as the long Taylor principle'' (p.~12), desirable ``independently of the forward lookingness of the economy and the degree of anchoring'' (p.~12); Section~\ref{sec:close} gives the checkable form in which he states it. %
The rung demotes the short-rate coefficient from a point to a sign, and moves the sign from a coefficient that cannot be identified from equilibrium data to a comovement those data can display.

A second current design evaluates the rule without a structural model. \citet{barnichon2023evaluating} measure the reaction function of US monetary policy, from the gold standard to the post-Volcker Federal Reserve, against the optimum in the space of impulse responses of the policy objectives to policy and non-policy shocks, and find little improvement in it until recent decades: a distance to the optimum stated conditional on the loss and on the identification of both sets of shocks. %

\subsection{What survives every rung}

Each relaxation in Table~\ref{tab:rungs} lowered the grade of what could be claimed. Within a maintained model the optimum was a point, stated with its author's condition attached. Across a named family it became a class of rules whose membership the family and the loss decided. Around a reference model it became a neighborhood with a chosen model at its center. Under misperceived natural rates, learning and the lower bound, it became directions and orderings, with magnitudes indexed by the noise, the gain and the shocks; on the current rung the one-for-one response gave way. No magnitude survived a single relaxation intact. The direction of the response to inflation survived them all: on every rung the surviving interest-rate rules raise the policy rate with inflation---``the most basic principle of monetary policy of leaning against the wind of inflation and output movements'' \citep[p.~851]{taylor2010simple}. %
The response to the output gap kept its sign until the last rung, where under optimal policy it can turn negative. Two signs remain unsettled: that of the adjustment to uncertainty, attenuation or aggressiveness, which belongs to what each exercise holds fixed, and that of the experimentation component, which is not fixed even within one framework.

This floor is what the rest of the survey inherits. An optimum computed on any rung takes its slope from the program of Section~\ref{sec:curve} and its account of the bank from that of Section~\ref{sec:rule}, and Section~\ref{sec:joint} showed that the two borrowings fail together; the optimal-policy tradition adds a twist, since a policy near the optimum removes from the data the covariation the slope estimate needs \citep{mcleay2020optimal}. The decision rules Section~\ref{sec:close} collects act on this floor, and so does the companion paper \citep{vansteenberghe2026tolerable}, whose bands for its policy object are adopted from published readings rather than identified and whose sign and ordering claims are confined to those that hold across the band. Within one model the tradition could state a point; across a named family, a band of rules; beyond any family it could name, only a sign---and the sign of the response to inflation is the one claim that every rung handed on to the next intact.

{\footnotesize
\setlength{\LTcapwidth}{\textwidth}
\begin{longtable}{>{\raggedright\arraybackslash}p{2.9cm}>{\raggedright\arraybackslash}p{2.8cm}>{\raggedright\arraybackslash}p{3.2cm}>{\raggedright\arraybackslash}p{2.8cm}>{\raggedright\arraybackslash}p{2.6cm}}
\caption{Seventy years of identification, V: the optimum---what each rung held fixed, what it showed, what the next rung relaxed, and the grade of claim that survived. Grades and sense tags as defined at the head of Section~\ref{sec:optimum}; `$\to$' reads `claimed $\to$ shown'; the pages in row four are those of \citet{levin1999robustness}. Table~\ref{tab:idhistory_rule} gives the estimation side of the rule.}\label{tab:rungs}\\
\toprule
Rung & What the rung held fixed & What the rung showed & What the next rung relaxed & What survives (sense claimed $\to$ shown) \\
\midrule
\endfirsthead
\multicolumn{5}{l}{\emph{Table~\ref{tab:rungs}, continued}}\\
\toprule
Rung & What the rung held fixed & What the rung showed & What the next rung relaxed & What survives (sense claimed $\to$ shown) \\
\midrule
\endhead
\bottomrule
\endlastfoot
Certainty equivalence, 1956--2003 \citep{simon1956dynamic,theil1957note} & a linear model, a quadratic loss, additive shocks, known parameters & the rule keeps its form; the loss absorbs the uncertainty & an uncertain multiplier (\citeauthor{brainard1967uncertainty}); several parameters (\citeauthor{chow1975analysis}); a restricted rule & the form (S3-point, in scope); attenuation for one multiplier independent of the shock (a sign) \\
\addlinespace[2pt]
Parameter uncertainty in a model, 1999--2001 \citep{estrella1999rethinking,rudebusch2001timid} & the model as the policymakers' own; the estimation covariance & negligible attenuation: the empty rung & an unrestricted VAR (\citeauthor{sack2000does}); uncertain persistence (\citeauthor{soderstrom2002monetary}) & a point, conditional on the model (S3-point) \\
\addlinespace[2pt]
Noisy states, 1999--2004 \citep{peersman1999taylor,swanson2004signal} & the model; known noise processes & the weight moves off the noisy argument & the natural rates as known quantities & the adjustment's direction (S3-point $\to$ sign) \\
\addlinespace[2pt]
Robust rules within a family, 1999 \citep{levin1999robustness} & parameters ``as known with certainty'' (p.~263, n.~1); four models & one model's optimum travels less well than simple rules; first-difference rules hold ``within this class of models'' (p.~294) & another family: indeterminacy, and money growth dominant under technology shocks (\citeauthor{christiano1999comment}) & a band (S3-point $\to$ band); one family away, a least-risk corner and the sign \\
\addlinespace[2pt]
The family widened and weighed, 2003--2010 \citep{levin2003robust,taylor2010simple} & five models or rival expectations; the weights on them & moderately inertial rules; a modest gap response, highly costly in no model & the loss weights; the model space itself & a band given family and loss; leaning against the wind \\
\addlinespace[2pt]
Robust control, 2000--2010 \citep{hansen2001robust,hansen2008robustness} & one reference model and a penalty & rules mostly more aggressive, as the doubt is written (\citeauthor{onatski2003modeling}) & the single reference model & a band around the reference, with its sign \\
\addlinespace[2pt]
Natural rates and learning, 2002--2008 \citep{orphanides2002robust,orphanides2008learning} & a given expectations scheme; then the learning gain & more on inflation; with misperceived natural rates, less on the perceived gap and more inertia & models not yet considered & directions (the counterpart of an S4-test across modeled expectations, passed by the sign) \\
\addlinespace[2pt]
Experimentation, 2000--2008 \citep{wieland2000monetary,svensson2008optimal} & a known linear form, submodels or modes & an incentive to experiment; gains later found small & robustness, which can mute or enhance it & smallness in the models tried; no fixed sign \\
\addlinespace[2pt]
The lower bound, 2000--2009 \citep{reifschneider2000three} & a linear model; normal shocks with the 1966--1995 residual covariance & costs rise nonlinearly as the target falls (\citeauthor{coenen2004price}) & the post-2008 shocks (\citeauthor{williams2009heeding}) & an ordering in target and shock size \\
\addlinespace[2pt]
The current rung, 2025 \citep{nakamura2025beyond,gali2025commentary} & the New Keynesian model; anchored expectations & less than one-for-one can be optimal; the inflation and gap coefficients can change sign & --- & a sign on the long real rate, after \citeauthor{gali2025commentary} (S3-point $\to$ sign) \\
\end{longtable}
}

\section{What the reconstruction licenses}
\label{sec:close}

\subsection{The register}

The two histories end in the same place. For the slope and for the rule's inflation response, each generation's estimate was identified conditional on an assumption the next generation relaxed, and Section~\ref{sec:joint} showed that the two programs fail together, each borrowing from the other the identification it lacks. A conclusion of that kind has a register, and the register is not a counsel of despair. Its rule is the one Section~\ref{sec:senses} introduced \citep[p.~12]{manski2013public}, %
which makes the ratchet that the tables record a general property of inference rather than a peculiarity of two literatures. Its method is \citeauthor{tamer2010partial}'s: rather than ``a binary answer that consists of whether the parameter is point identified'', the approach ``characterizes the informational content of various assumptions by providing a menu of estimates, each based on different sets of assumptions, some of which are plausible and some of which are not'' \citep[p.~167]{tamer2010partial}. %
Its object is \citeauthor{lewbel2019identification}'s: when the model and the population do not determine the parameter uniquely, the report is ``the tightest bounds possible on $\theta$, given $M$ and [what is knowable from the data]'' \citep[p.~877]{lewbel2019identification}. %
Its communication rule is Manski's as well: exact predictions and point estimates reported with incredible certitude ``should not be a general practice'' \citep[p.~216]{manski2020lure}. %
The register is older than its name. \citeauthor{leamer1983lets}'s fragility test, quoted in Section~\ref{sec:senses} \citep[p.~38]{leamer1983lets}, is the same rule stated as a warning, %
and \citeauthor{sims1980macroeconomics}'s charge that the objections to the large models ``make it unlikely that macroeconomic models are in fact over identified, as the existing statistical theory usually assumes'' \citep[abstract, p.~1]{sims1980macroeconomics} is the same rule turned on the identifying restrictions of the day.

\subsection{What it licenses for the two coefficients}

The register licenses three things, and the reconstruction supplies the material for each.

First, a set-valued report. Neither coefficient can be point-identified from the equilibrium histories without an assumption the next generation found incredible, and what can honestly be reported for each is the range its credible assumptions leave open. The exhibits are in the tables: the identification-robust confidence interval of \citet{kleibergen2009weak} for the price-stickiness parameter behind the New Keynesian slope, ``though unbounded from above, \ldots\ still informative'' (p.~294), %
and the post-1979 sets of \citet{mavroeidis2010monetary} for the rule, in which ``the data are also consistent with the opposite view that policy remained inactive after 1979'' (p.~491). %
In the full-information writings the same fact appears as a likelihood with a ridge, nearly flat in some directions \citep{schorfheide2013estimation}, %
along which the prior is not updated. The three treatments of that flat direction disagree in a way the reader should see rather than have settled for him. \citet{moon2012bayesian} show that ``asymptotically Bayesian highest-posterior-density sets exclude parts of the estimated identified set'' (p.~755), %
so that in the flat direction the narrowness of a credible set is the prior's work. They recommend ``reporting estimates of the identified set and information about the conditional prior along with Bayesian credible sets'' (p.~755). \citet{baumeister2015sign} accept the diagnosis and prescribe the opposite cure, an explicit and defended prior, since ``the influence of these priors does not vanish asymptotically'' (abstract); %
\citet{giacomini2021robust} refuse the single prior, with a class of priors that ``removes the source of the disagreement: the need to specify an unrevisable prior for the structural parameter given the reduced-form parameter'' (p.~1519). %
The choice among the three is the reader's; the tag S3-P of Section~\ref{sec:senses} marks the rows on which a prior has done the work.

Second, sign-and-order claims that hold across every credible design. This is the survey's reading of its tables, not a proposition any source states in these words: the slope is positive and small in every design that does not assume it away, and the rule's response to inflation is positive in every reading of the post-1979 data, while its magnitude is not. A claim of this form is what the key of Section~\ref{sec:senses} leaves standing when the sense falls from S3 to S5---the ordering survives the demotion; the point does not. The same form appears from inside the New Keynesian model in \citet{gali2025commentary}, for whom, when the rule is written around the path the bank seeks to implement, the Taylor principle ``applies only to off-equilibrium paths that are not observed by an econometrician'' (p.~7), %
while ``a positive comovement between the long real rate and inflation is a condition that a desirable policy in a variety of environments must satisfy, and one that can be easily checked'' (p.~17): %
a sign condition that equilibrium data can check, offered in place of a coefficient they cannot deliver.

Third, a decision rule that acts on the direction and observes the outcome. The classic formal statement is the attenuation \citet{brainard1967uncertainty} derived for an uncertain multiplier, which Blinder kept as a practice without a theorem: ``There is no Brainard conservatism theorem. Yet I still believe what I wrote'' \citep[p.~35]{blinder2006monetary}. \citeauthor{greenspan2004risk}'s risk management applies the rule to a distribution the model does not pin down, though as insurance it can call for a larger move rather than a smaller one (p.~37): ``Policymakers often have to act, or choose not to act, even though we may not fully understand the full range of possible outcomes'' \citep[p.~38]{greenspan2004risk}. %
\citet{orphanides2002robust} gave it an operational form, difference rules ``which do not require knowledge of the natural rates of interest and unemployment and are consequently immune to likely misperceptions in these concepts'' (p.~66), a rule that responds to what is observed rather than to a level the data cannot pin down. Hansen and Sargent's robust decision maker, who ranks decisions by their worst case in a neighborhood of a reference model \citep{hansen2001robust,hansen2008robustness}, keeps a model that \citeauthor{manski2004statistical}'s planner does without: the planner who cannot place a subjective distribution on the unknowns chooses by minimax regret \citep{manski2004statistical}. %
\citet{christensen2025optimal} derive decision rules for choices whose payoffs depend on a partially identified parameter: minimax where partial identification leaves the parameter ambiguous, average-risk minimization over the point-identified parameter that restricts it. %
And \citet{kocherlakota2019practical} states the rule for the decision a reader of this survey actually faces, the next move within a regime: ``The regression function (of past payoffs on past choices) is still a sufficient statistic for the current policy decision'' (p.~30), %
a rule that needs no structural model and that its author does not extend to ``a once-and-for-all change in an overarching policy regime'' (p.~29). %
None of the three licenses a point, and none requires one. Each acts on a sign, an ordering, or a regression of outcomes on choices, and each observes what follows.

\subsection{The equivalence statements that bracket the seventy years}
\label{sec:close-oe}

The register also settles how the observational-equivalence results at the two ends of the reconstruction are to be read. \citet{sargent1976observational} opened the sequence: the natural- and unnatural-rate writings of the output equation are observationally equivalent (Section~\ref{sec:senses}), so that the fit of either is not a test against the other. \citet{mccallum1979observational} argued that the equivalence can be broken. If the classical supply function is fully specified, so that every relevant state variable enters the equation, only the current surprise moves output---in \citeauthor{mishkin1983rational}'s summary of the argument, ``if all the state variables are included in the MRE output or unemployment equation, then the theory does imply that $N=0$'' \citep[p.~22]{mishkin1983rational}---and the two writings then differ. That is what breaks the equivalence; its price is stated by the same monograph and by the discussants of the day, and it is a maintained restriction. The equivalence ``is overcome, parameters are identified, and tests of the MRE model are feasible'' only under one of two conditions, a known lag length (zero in the efficient-markets case) or an exclusion (pp.~15, 31), and on the first ``since relevant state variables are almost surely excluded from estimated MRE equations, the lag length is not known'' (p.~22); ``a priori restrictions, by definition, are not tested'', yet ``any estimation requires some identifying restrictions, some maintained hypothesis'' \citep[p.~52]{blinder1980comment}; a reduced form for prices in which ``the same variables appear in each'' model, market-clearing or with gradual price adjustment, cannot say which one it tests \citep[p.~60]{gordon1980comment}; and \citet{minford1979classical} showed that other versions of the classical supply hypothesis do not deliver the restriction. Sargent's reply to McCallum \citep{sargent1979causality} completes the exchange. %
What the profession retained is the register's own conclusion: the equivalence is broken by an assumption, and the assumption is the price.

The later statements are the same transaction at other prices. On the rule, \citet{rudebusch2002term} showed that quarterly inertia ``would imply a large amount of forecastable variation in interest rates at horizons of more than 3 months, which is contradicted by evidence from the term structure of interest rates'', so that ``the illusion of monetary policy inertia evident in the estimated policy rules likely reflects the persistent shocks that central banks face'' (p.~1161): %
two writings the rule regression cannot separate, separated by an off-model check. At the system level, \citet{canova2009back} found the identification problems quoted in Section~\ref{sec:senses} to be ``typically produced by an ill-behaved mapping between the structural parameters and the coefficients of the solution'' (p.~431). %
\citet{cochrane2023fiscal} gives the general form quoted in Section~\ref{sec:senses}, and its reason: ``The regimes differ in how we imagine the government behaves away from equilibrium'', and ``We can't observe that behavior in data drawn from the equilibrium'' (p.~488). %
\citet{chahrour2018news} proved an equivalence of the same form on the expectations side (Section~\ref{sec:senses}), there without a price: ``neither representation requires stronger modeling assumptions'' (p.~1703). %
In every case the object in dispute is one the equilibrium path does not display, and more data drawn from that path will not display it either.

\paragraph{The motivating case.}
Why a reader outside monetary economics should care is best seen on one policy object, and the tolerance dial of the companion paper \citep{vansteenberghe2026tolerable} is the case this survey was written for. The dial measures how much of a supply-driven inflation a central bank chooses to look through, and every ingredient of its optimum rests on one of the two coefficients whose histories the tables record. The slope enters twice: through the intensity $\lstar=1/(1+\ep\kappa)$, which fixes how far it is optimal to accommodate a given supply impulse, and through the persistence of inflation, which a curve near flat makes uninformative about how much was tolerated. The rule's response enters once, because a tolerated inflation can only be read as a softening relative to what the rule would otherwise have done. The slope the dial needs is the aggregate price-side slope on the activity variable the rule reads, which Section~\ref{sec:curve} showed to be one model step beyond any object the credible modern designs deliver, and the response it needs is the one Section~\ref{sec:rule} showed to be contested even in principle. The companion then adds two objects neither program ever required: the share of an impulse that is supply rather than demand, latent at the moment the stance must be chosen, and the menu of responses the bank did not choose, a counterfactual that no record of the equilibrium path contains. Nor is the public's monitor better placed than the econometrician: the canonical scheme for detecting an implicit target, \citeauthor{svensson1997inflation}'s inflation-forecast targeting, has the public compare the observed instrument with what a revealed model prescribes, so its benchmark inherits the coefficients the tables show cannot be pinned---a difficulty its author raised as a possible objection, that ``the lack of knowledge, and resulting disagreement, about the appropriate macroeconomic model \ldots\ may be so substantial as to make both the implementation and monitoring of the target rule proposed here too difficult'' \citep[p.~1138]{svensson1997inflation}. %
The companion reads a recent credibility model, that of \citet{afrouzi2026monetary}, as one more instance: regimes that coincide in the steady state that the calibration targets (p.~2444) and separate in the inflation that follows a shock, where ``a small change in the labor wedge has a sizable impact on long-run inflation'' (p.~2445). The reconstruction therefore sets a floor under the composite rather than a background for it: a function of two inputs that seventy years of designs could not pin down will not be pinned down by a further estimator run on the same equilibrium record, and the failure is inherited in the correlated form Section~\ref{sec:joint} described. What the register licenses instead is what the companion does: report the range that the data and the readings one is prepared to defend leave open; say which ingredient contributes how much of its width; and build both the decision and the way it is explained for that range rather than for a point the data will not supply. The bands the companion reports are adopted from credible readings, not identified, and the reading of them this survey supports is the one its tables support for the inputs---a direction, an ordering, and a range. %

\newpage
\bibliographystyle{aea}
\bibliography{literature}

\end{document}